\documentclass[sigconf]{acmart}
\usepackage{enumitem}
\usepackage{multirow}
\usepackage{subfigure}
\usepackage[ruled,vlined]{algorithm2e}

\newtheorem{definition}{Definition}

\AtBeginDocument{%
  \providecommand\BibTeX{{%
    \normalfont B\kern-0.5em{\scshape i\kern-0.25em b}\kern-0.8em\TeX}}}

\copyrightyear{2022}
\acmYear{2022}
\setcopyright{acmcopyright}
\acmConference[SIGIR '22] {Proceedings of the 45th International ACM SIGIR Conference on Research and Development in Information Retrieval}{July 11--15, 2022}{Madrid, Spain.}
\acmBooktitle{Proceedings of the 45th International ACM SIGIR Conference on Research and Development in Information Retrieval (SIGIR '22), July 11--15, 2022, Madrid, Spain}
\acmPrice{15.00}
\acmISBN{978-1-4503-8732-3/22/07}
\acmDOI{10.1145/3477495.3532014}

\begin{document}
\fancyhead{}
\title{Less is More: Reweighting Important Spectral Graph Features for Recommendation}

\author{Shaowen Peng}
\email{swpeng95@gmail.com}
\affiliation{%
  \institution{Kyoto University}
  \city{Kyoto}
  \country{Japan}
}

\author{Kazunari Sugiyama}
\email{kaz.sugiyama@i.kyoto-u.ac.jp}
\affiliation{%
  \institution{Kyoto University}
  \city{Kyoto}
  \country{Japan}
}

\author{Tsunenori Mine}
\email{mine@m.ait.kyushu-u.ac.jp}
\affiliation{%
  \institution{Kyushu University}
  \city{Fukuoka}
  \country{Japan}
}

\begin{abstract}
As much as Graph Convolutional Networks (GCNs) have shown tremendous success in recommender systems and collaborative filtering (CF), the mechanism of how they, especially the core components (\textit{i.e.,} neighborhood aggregation) contribute to recommendation has not been well studied. To unveil the effectiveness of GCNs for recommendation, we first analyze them in a spectral perspective and discover two important findings: (1) only a small portion of spectral graph features that emphasize the neighborhood smoothness and difference contribute to the recommendation accuracy, whereas most graph information can be considered as noise that even reduces the performance, and (2) repetition of the neighborhood aggregation emphasizes smoothed features and filters out noise information in an ineffective way. Based on the two findings above, we propose a new GCN learning scheme for recommendation by replacing neihgborhood aggregation with a simple yet effective Graph Denoising Encoder (GDE), which acts as a band pass filter to capture important graph features. We show that our proposed method alleviates the over-smoothing and is comparable to an indefinite-layer GCN that can take any-hop neighborhood into consideration. Finally, we dynamically adjust the gradients over the negative samples to expedite model training without introducing additional complexity. Extensive experiments on five real-world datasets show that our proposed method not only outperforms state-of-the-arts but also achieves 12x speedup over LightGCN. 

\end{abstract}



\maketitle

\section{Introduction}
Recommender systems have been playing an important role in people's daily life by predicting items the user may be interested in based on the analysis of users' historical records, such as user-item interactions, user reviews, demographic data, etc. Collaborative Filtering (CF), which focuses on digging out the user preference from past user-item interactions, is a fundamental task for recommendation. A common paradigm for CF is to characterize users and items as learnable vectors in a latent space and optimize based on user-item interactions. Matrix factorization (MF) \cite{koren2009matrix} is one of the most widely used embedding-based methods, which estimates the ratings as the inner product between user and item latent vectors. To overcome the drawback of MF that simply exploits a linear function to model complex user behavior, subsequent works exploit advanced algorithms such as perceptrons \cite{he2017neural,covington2016deep}, recurrent neural networks \cite{wu2017recurrent,hidasi2015session}, memory networks \cite{ebesu2018collaborative}, attention mechanisms \cite{chen2017attentive}, transformer \cite{sun2019bert4rec} to model non-linear user-item relations.\par 

However, the unavailability of capturing the higher-order signals limits the performance of the aforementioned methods due to the data sparsity. Graph convolutional networks (GCNs) \cite{defferrard2016convolutional,kipf2017semi} have attracted much attention and have shown great potential in various fields including social network analysis \cite{wu2019neural,fan2019graph} and recommender systems \cite{berg2017graph,wang2019neural}. The core idea of GCNs is to augment node representations with (higher-order) neighborhood. Much effort has been devoted to adapt GCNs \cite{kipf2017semi} to CF. For instance, NGCF \cite{wang2019neural} is inspired and inherits the components from vanilla GCN \cite{kipf2017semi}; EGLN \cite{yang2021enhanced} learns an adaptive user-item graph structure to predict potential positive preference. Some research efforts have been made to simplify GCN-based CF methods by removing the non-linearity and embedding transformation \cite{he2020lightgcn, chen2020revisiting}. However, we notice that the mechanism of how GCNs contribute to recommendation has not been well studied. To this end, our work focuses on the core component (i.e, neighborhood aggregation) and aims to investigate the following research questions:
\begin{itemize}[leftmargin=11pt]
\vspace{-0.06cm}
\item What graph information matters for recommendation?

\item How neighborhood aggregation helps recommendation and why repeating it achieves better accuracy?

\item Is there a more effective and efficient design to replace neighborhood aggregation?
\vspace{-0.06cm}
\end{itemize}
To answer the aforementioned questions, we review GCNs from a spectral perspective in Section 3.1. Specifically, we decompose adjacency matrix into spectral features and discover two important findings: (1) we identify neighborhood smoothness and difference which significantly affect the recommendation accuracy while only account for a small portion of the spectral features, whereas most graph information has no positive effect that can be considered noise added on the graph; (2) stacking layers in GCNs tends to emphasize graph smoothness and depress other information. Based on the two findings above, we unveil the ineffectiveness of neighborhood-aggregation and replace it with our proposed Graph Denoising Decoder (GDE) in Section 3.2, which only keeps the important graph features to model the graph smoothness and difference without stacking layers. Our proposed GDE significantly simplifies existing GCN-based CF methods and reduces the running time. The contributions of our work are summarized as follows:

\begin{itemize}[leftmargin=10pt]
\item We unveil the effectiveness of Graph Convolutional Networks (GCNs) for recommendation in a spectral perspective, and shed light on the ineffectiveness of the existing design (\textit{i.e.,} neighborhood aggregation), which provides theoretical and empirical support for our proposed method.   
 
\item Compared with existing work that stacks many layers to capture higher-order neighborhood, our proposed GDE is built in a different way by directly capturing the important spectral graph features to emphasize the neighborhood smoothness and difference, which is equipped with a simple yet effective architecture that can incorporate neighborhood signals from any hops. 

\item We propose to dynamically adjust the magnitude of gradients over the negative samples to tackle the slow convergence issue on GCN-based CF methods, resulting in further improvement and helping expedite the model training.

\item Extensive experiments on five real-world datasets not only show our proposed method outperforms state-of-the-arts under extreme data sparsity with less running time but also demonstrate the effectiveness of our proposed designs. 
\end{itemize}

\section{Preliminaries}
\subsection{Graph Convolutional Network for CF}
We first summarize the common GCN paradigm for CF. Given an interaction matrix $\mathbf{R}=\{0, 1\}^{M\times N}$ with $M$ users and $N$ items consisting of observed $\mathbf{R}^+$ and unobserved interactions $\mathbf{R}^{\mbox{-}}$, we define a bipartite graph $\mathcal{G}=(\mathcal{V},\mathcal{E})$ where the node set $\mathcal{V}$ contains all users and items, $\mathcal{E}=\mathbf{R}^+$ are the edge set. Each user and item is considered as a node of $\mathcal{G}$ and is characterized as a learnable embedding vector $\mathbf{e}_u\in \mathbb{R}^d$ ($\mathbf{e}_i\in \mathbb{R}^d$); by stacking them together we have an embedding matrix $\mathbf{E}\in\mathbb{R}^{(M+N)\times d}$. The goal is to estimate the unobserved interactions $\mathbf{R}^{\mbox{-}}$, by learning an interaction function given as follows:
\begin{equation}
f(\mathbf{R}^{\mbox{-}}|\mathcal{G},\mathbf{R}^+,\mathbf{ \Theta}):\mathcal{V}_u\times \mathcal{V}_i\rightarrow \mathbb{R}^+, 
\end{equation} 
where ${\mathbf \Theta}$ denotes model parameters. Here, $f(\cdot)$ corresponds to a specific GCN model. The matrix and node form of the updating rule are generally formulated as follows:

\begin{equation}
\begin{aligned}
&\mathbf{H}^{(k+1)}=\sigma\left( \mathbf{\hat{A}} \mathbf{H}^{(k)} \mathbf{W}^{(k+1)} \right),\\
&\mathbf{h}^{(k+1)}_{u}=\sigma\left(\sum_{i\in\mathcal{N}_u} \frac{1}{\sqrt{d_u+1}\sqrt{d_i+1}}\mathbf{h}^{(k)}_{i} \mathbf{W}^{(k+1)}\right),
\end{aligned}
\label{gcn}
\end{equation}
where $\sigma(\cdot)$ is a non-linear activation function; $\mathbf{W}^{(k+1)}$ are the weight matrix at $(k+1)$-th layer, $d_u$ and $d_i$ are the node degree for $u$ and $i$, respectively; $\mathcal{N}_u$ are nodes directly connected to $u$, $\mathbf{\hat{A}}=\mathbf{\tilde{D}}^{-\frac{1}{2}}\mathbf{\tilde{A}}\mathbf{\tilde{D}}^{-\frac{1}{2}}$, and $\mathbf{\tilde{D}}=\mathbf{D}+\mathbf{I}$, $\mathbf{\tilde{A}}=\mathbf{A}+\mathbf{I}$, where $\mathbf{A}$, $\mathbf{D}$ and $\mathbf{I}$ are the adjacency matrix, diagonal degree matrix and identity matrix, respectively. The node embedding is updated by aggregating the neighborhood's current embedding through adjacency matrix starting from the initial state $\mathbf{h}_u^{(0)}=\mathbf{e}_u$. Some works simplify the updating rule by removing the activation function or weight matrix and gain further improvements \cite{chen2020revisiting,he2020lightgcn}. The final node embeddings are generated from the previous embeddings via a pooling function: 
\begin{equation}
\mathbf{o}_u=\mathbf{pooling}\left(\mathbf{h}^{(0)}_u,\cdots,\mathbf{h}^{(K)}_u\right).
\label{framework}
\end{equation}
Common pooling functions are sum, concatenate; some works also output the last layer as the final embeddings. Finally, an interaction between a user and an item is estimated as follows:
\begin{equation}
\mathbf{\hat{r}}_{ui}=\mathbf{o}_u^T\mathbf{o}_i.
\end{equation} 

\subsection{Graph Signal Processing}
We introduce an important definition from graph signal processing. Given a graph signal $\mathbf{s}$, its variation on the graph is defined as:
\begin{equation}
\|\mathbf{s}-\mathbf{\hat{A}}\mathbf{s}\|.
\end{equation}
\label{variation}
The variation of a signal on a graph measures the difference between the signal at each node and at its neighborhood \cite{sandryhaila2014discrete}. Intuitively, a signal with small variation   implies the smoothness between each node and its neighborhood, whereas a signal with large variation stresses the difference between them.

\begin{definition}
Given eigenvalue $\lambda_t$ and eigenvector $\mathbf{v}_t$ of $\mathbf{\hat{A}}$, the variation of an eigenvector on the graph is $\|\mathbf{v}_t-\mathbf{\hat{A}}\mathbf{v}_t\|=1-\lambda_t$.
\end{definition}
Where $\lambda_t\in(-1,1]$ \cite{li2018deeper}. According to eigendecomposition $\mathbf{\hat{A}}=\mathbf{V}diag(\lambda_t)\mathbf{V}^T=\sum_t \lambda_t \mathbf{v}_t \mathbf{v}_t^T$, the graph information is made up of orthogonal spectral features. Through Definition 1, we can classify these features based on their variations: $\mathbf{v}_t$ with larger eigenvalue is smoother (smaller variation), while the feature with smaller eigenvalue is rougher (larger variation). In this work, we focus on investigating how spectral features with different variations affect recommendation accuracy.

\begin{figure*} \centering 
\subfigure[CiteUlike on vanilla GCN. ] {  
\includegraphics[width=0.47\columnwidth]{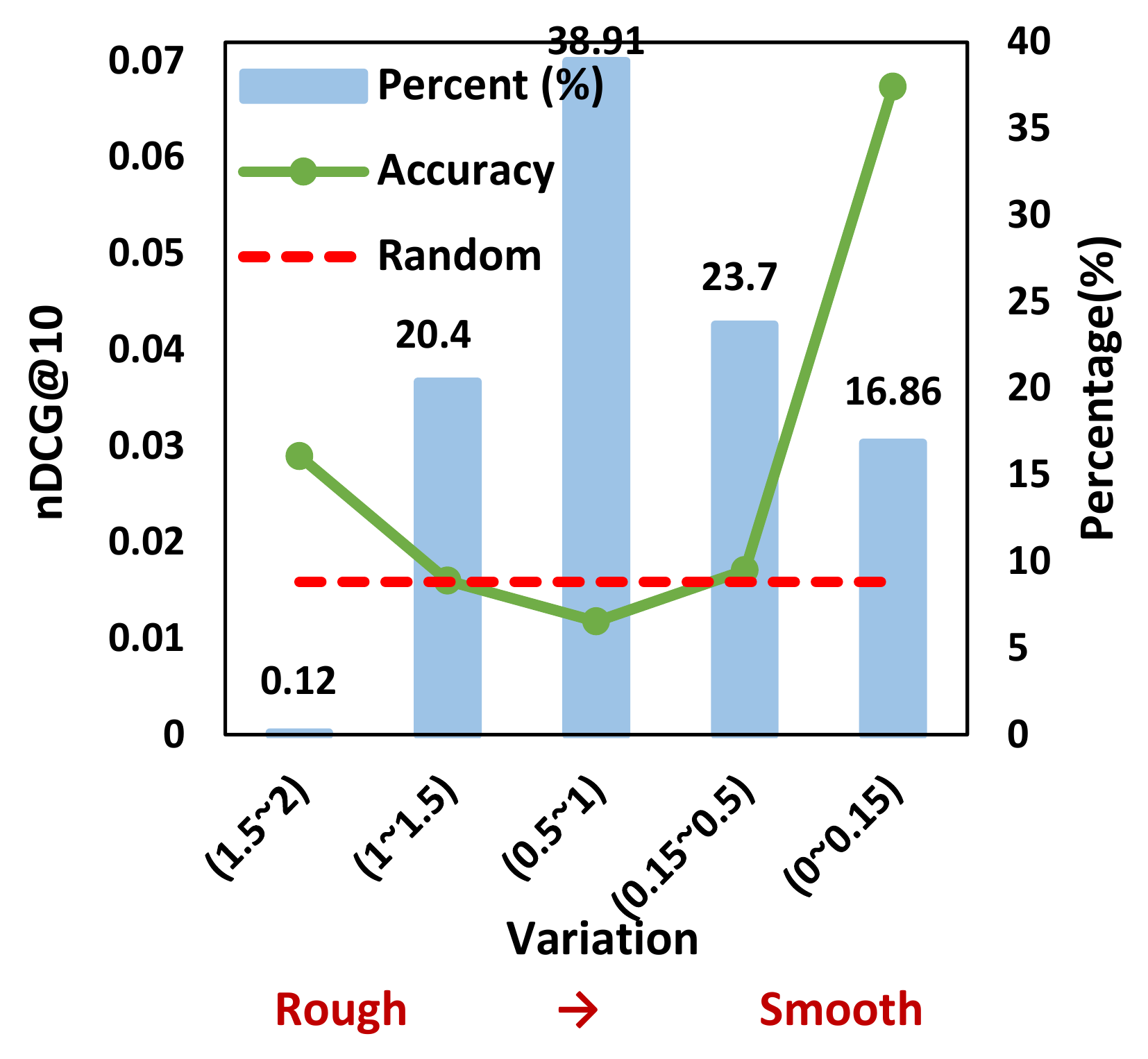} 
}
\hspace{0.05cm}
\subfigure[MovieLens on vanilla GCN. ] {  
\includegraphics[width=0.47\columnwidth]{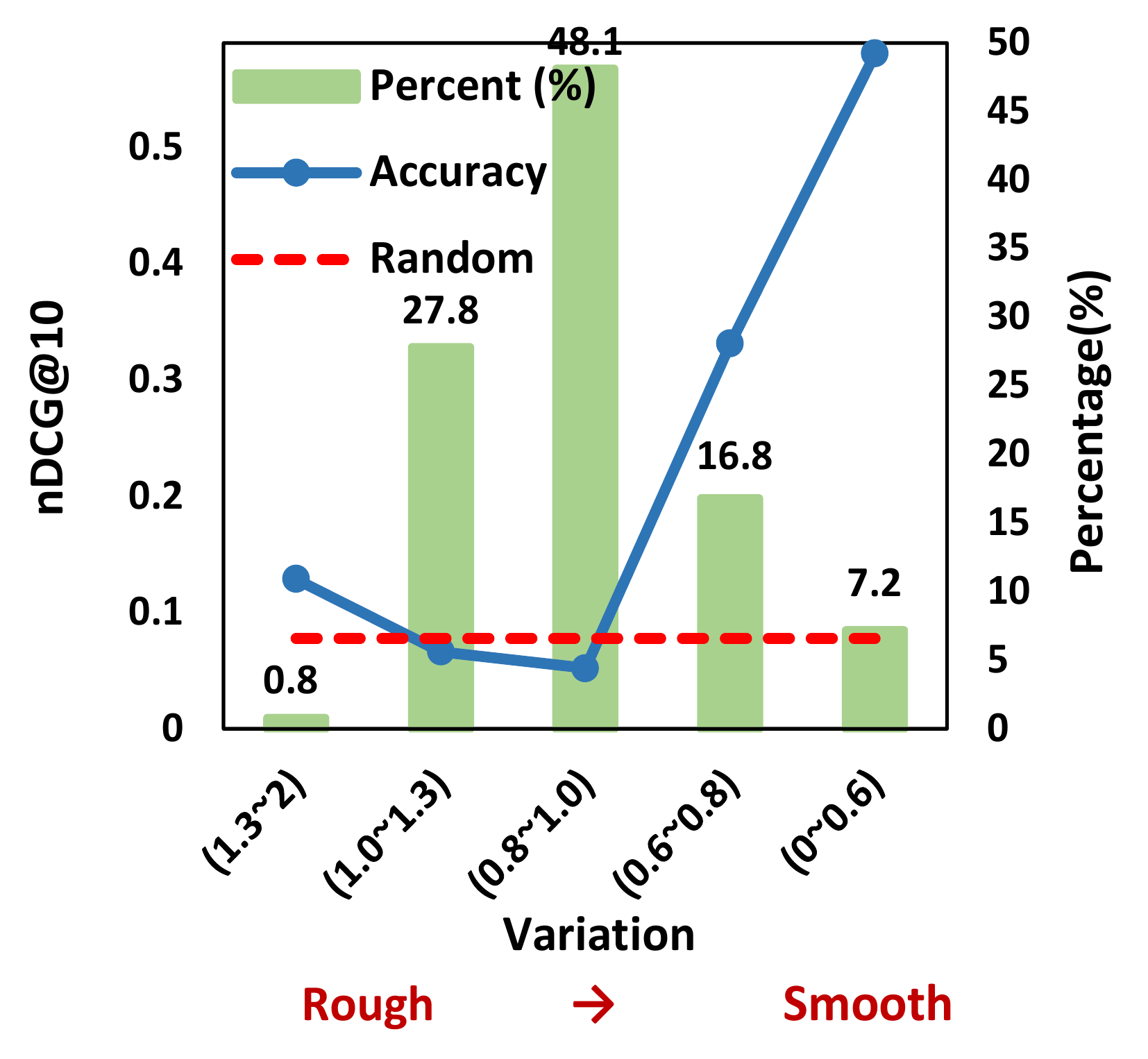} 
}
\subfigure[CiteULike on LightGCN. ] {  
\includegraphics[width=0.47\columnwidth]{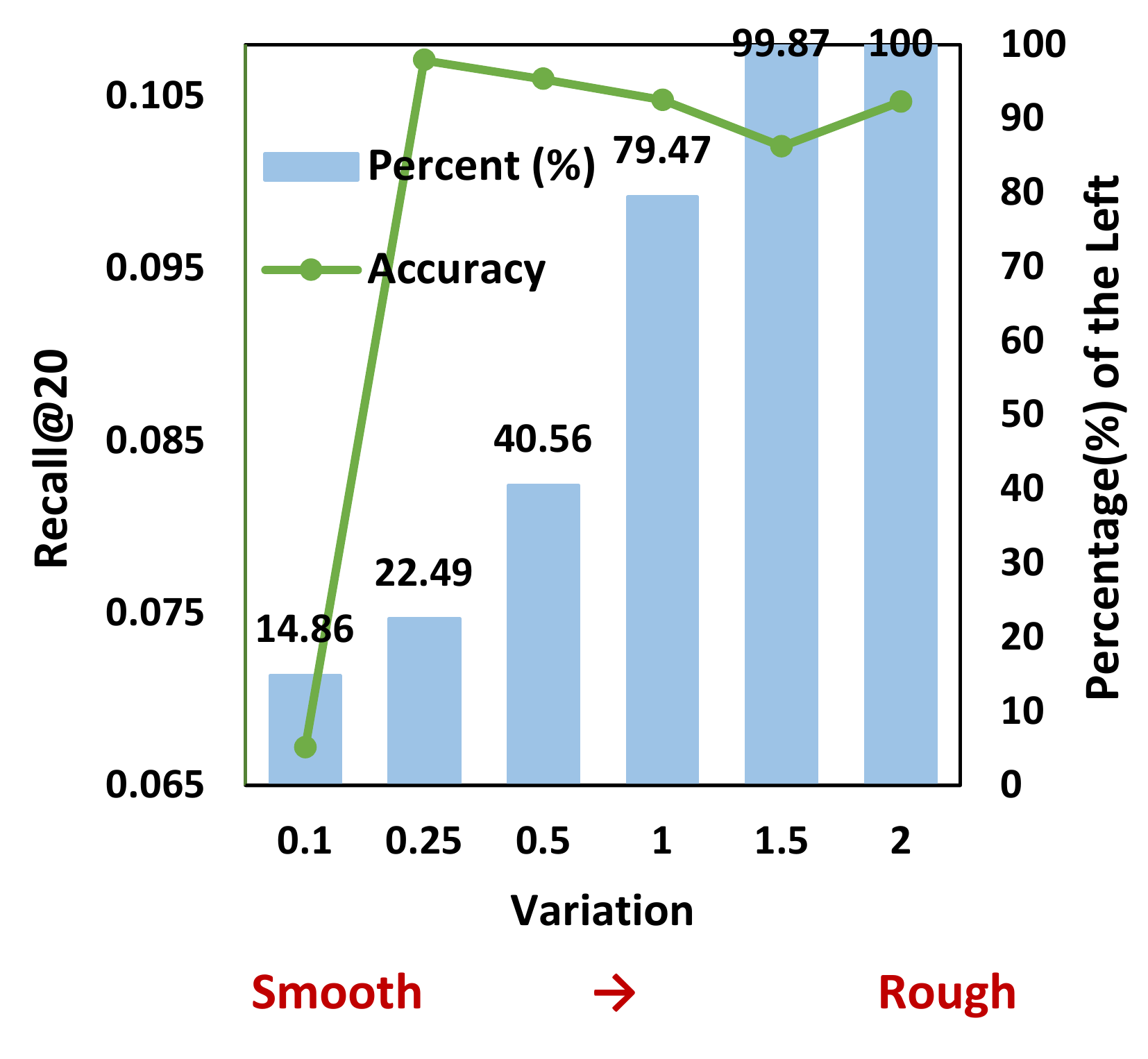} 
}
\hspace{0.05cm}
\subfigure[MovieLens on LightGCN. ] {  
\includegraphics[width=0.47\columnwidth]{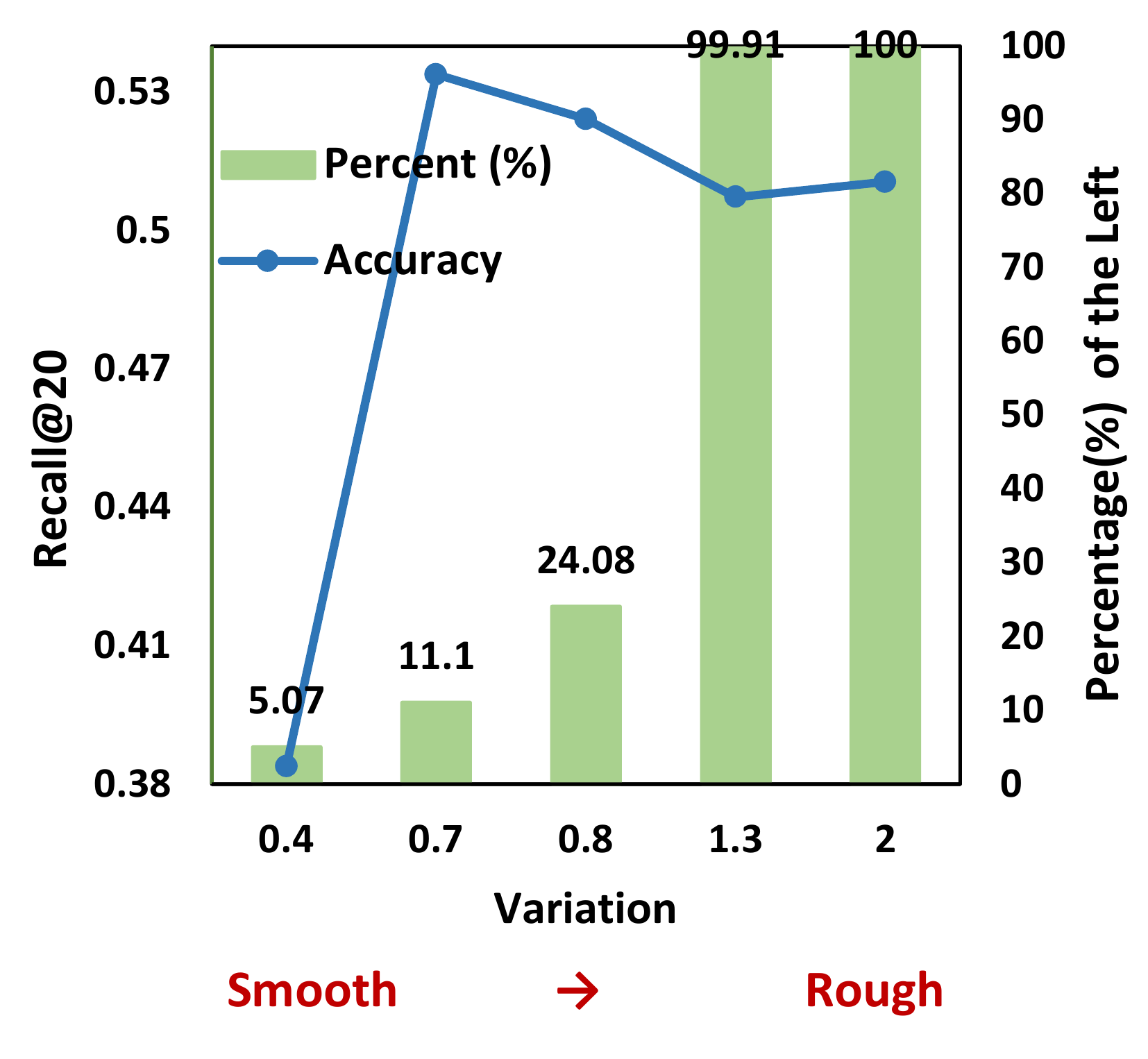} 
}
\vspace{-0.3cm}
\caption{In (a) and (b), we partition spectral features into different groups based on their variations, it illustrates the percentage of features in different groups and how they contribute to the accuracy which is tested on vanilla GCN; we use a randomly initialized adjacency matrix for 'random' as the benchmark. (c) and (d) show the accuracy where the features with variation $\geq x$ are removed (\textit{e.g.,} the result on $x=2$ is obtained on the original graph containing all features) on LightGCN.}  
\label{signal_importance}
\end{figure*}

\section{Methodology}
\subsection{Recap in A Spectral Perspective}
As recent works \cite{chen2020revisiting,he2020lightgcn} show that GCNs perform better without non-linear activation functions and transformation for CF, the effectiveness of GCNs lies in neighborhood aggregation, which is implemented as the multiplication of adjacency matrix. Following Definition 1, we study how neighborhood aggregation affects and contributes to recommendation accuracy.   

\subsubsection{The Importance of Different Spectral Graph Features}
We first define a cropped adjacency matrix as:
\begin{equation}
\mathbf{\hat{A}}'=\sum_t \mathcal{M}(\lambda_t) \mathbf{v}_t \mathbf{v}_t^T,
\label{crop}
\end{equation}  
where $\mathcal{M}(\lambda_t)=\{0, \lambda_t\}$ is a binary value function. We only keep the tested features ($\mathcal{M}(\lambda_t)=\lambda_t$) and remove others ($\mathcal{M}(\lambda_t)=0$) to verify the effectiveness of different spectral features for CF. We replace $\mathbf{\hat{A}}$ with Equation (\ref{crop}) and apply it to two classic models: vanilla GCN \cite{kipf2017semi} and LightGCN \cite{he2020lightgcn} that have been extensively adopted as baselines for CF, and conduct experiments on two datasets: CiteULike (sparse) and MovieLens-1M (dense) (see Table \ref{datasets} for details). Figure \ref{signal_importance} shows the results.\par

In Figure \ref{signal_importance} (a) and (b), the accuracy is mainly contributed by a small portion of features that are rather smoothed or rough, while most features concentrated in the middle area (83\% on CiteULike and 76\% on MovieLens) which are not so rough or smoothed contribute as little as the randomly initialized signals to the accuracy and are barely useful. If we ignore the noisy features, it is obvious that the smoother features tend to outweigh the rougher features. In Figure \ref{signal_importance} (c) and (d), we evaluate the importance of certain features by observing how accuracy changes after removing them. Slight and significant drops in accuracy are identified right after removing the top rough and smoothed features, respectively, while removing other features even results in an improvement.\par

The above observations raise a question: why the accuracy is mainly contributed by the top smooth and rough features? According to the definition, the two kinds of features actually represent the tendency of user behaviour: homophily and heterophily, that a user tends to interact with both similar others (\textit{i.e.,} in the neighborhood) and different others (\textit{i.e.,} not related to the user) \cite{ramazi2018homophily}, whereas most features that are not either rather smoothed or rough are less helpful to emphasize the two effects, thus contribute less to the accuracy. To summarize the analysis:

\begin{itemize}[leftmargin=10pt]
\item Only a small portion of features that rather smoothed or rough are truly helpful for the recommendation.

\item Smoothed features (homophily) outweigh the rough features (heterophily) for recommendation.
\end{itemize}
Note that this finding only applies to recommendation tasks, because the importance of different spectral features varies on tasks according to their data characteristics \cite{wang2020high,fagcn2021}.

\subsubsection{Analysis and Limitations of Existing Work}
Our analysis is based on LightGCN \cite{he2020lightgcn} which only keeps the essential component (\textit{i.e.,} neigbhrhood aggregation), and we can rewrite it as: 
\begin{equation}
\mathbf{O}=\sum_{k=0}^{K}\alpha_k\mathbf{H}^{(k)}=\sum_{k=0}^{K}\frac{\mathbf{\hat{A}}^k}{K+1}\mathbf{E}=\left( \sum_t \left( \sum_{k=0}^K  \frac{\lambda_t^k }{K+1 }\right) \mathbf{v}_t \mathbf{v}_t^T\right) \mathbf{E}.
\label{graph_filter}
\end{equation}
\begin{figure} \centering 
\vspace{-0.7cm}
\subfigure[on LightGCN.] {  
\includegraphics[width=0.46\columnwidth]{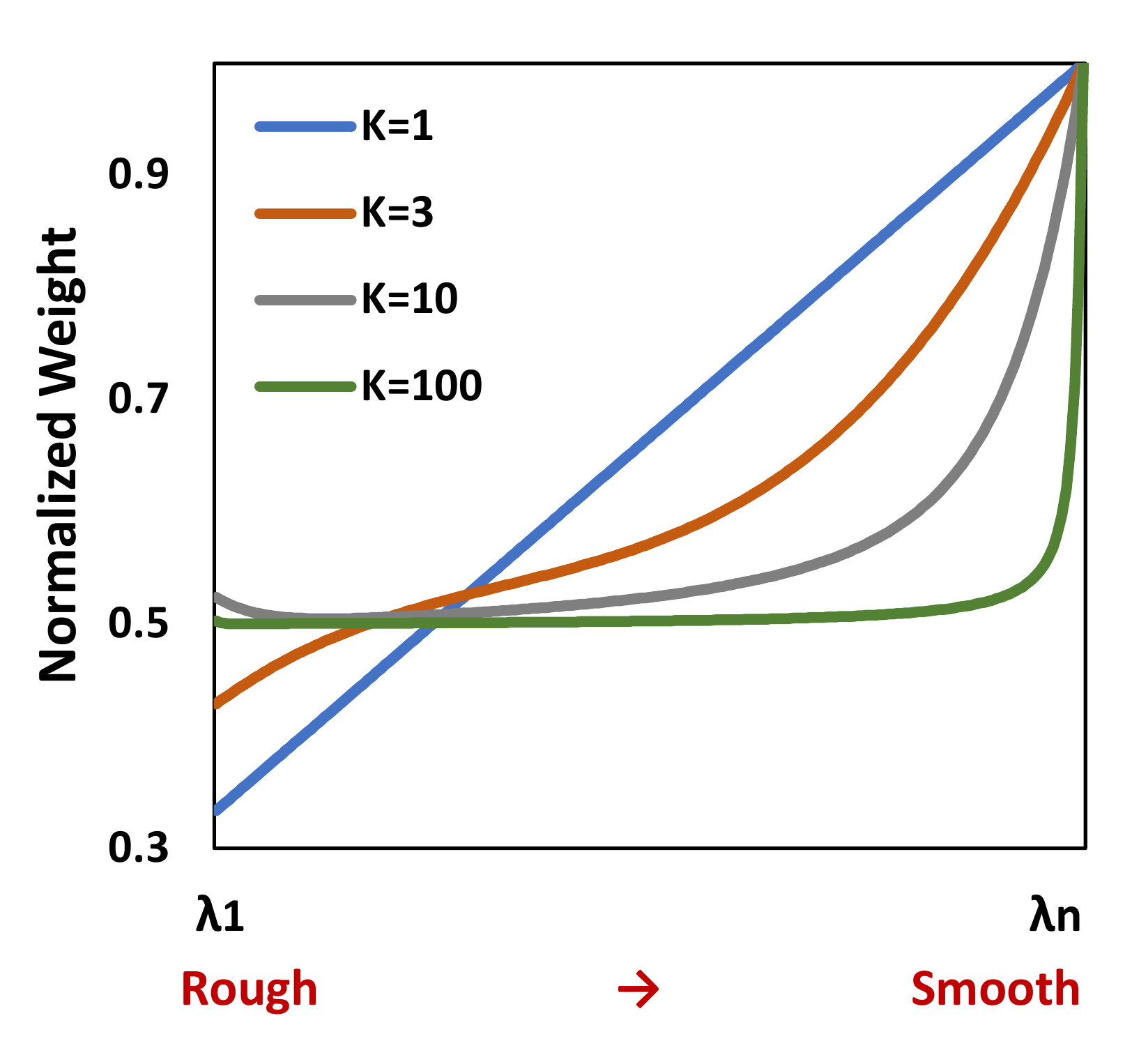} 
}  
\hspace{-0cm}
\subfigure[An ideal filter. ] {  
\includegraphics[width=0.46\columnwidth]{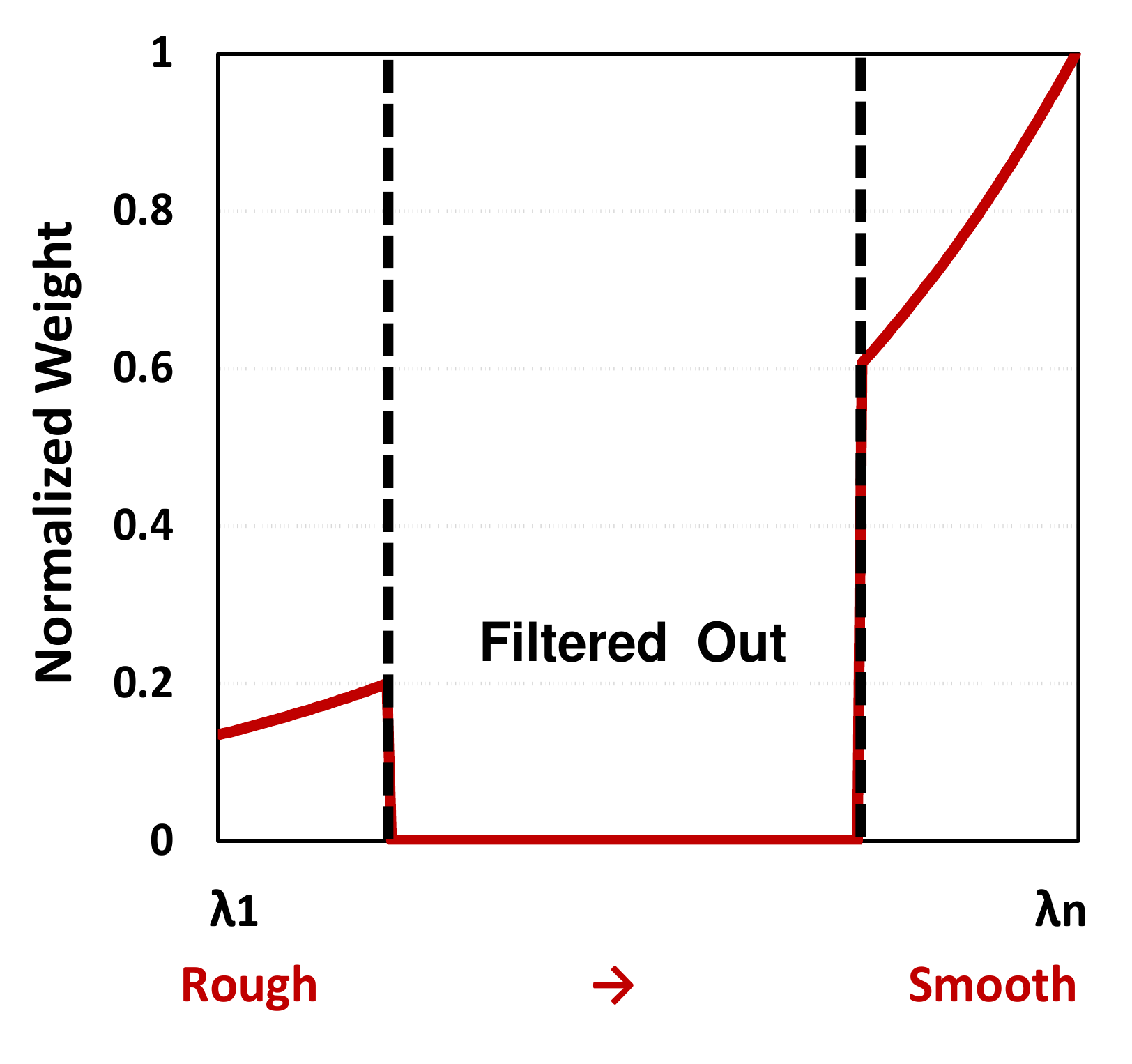} 
}
\vspace{-0.2cm}
\caption{The normalized weights for distinct graph features.}  
\label{weight}
\vspace{-0.5cm}
\end{figure}
We can see each spectral feature is weighted by a polynomial filter $\sum_{k=0}^K \alpha_k \lambda_t^k $ ($\alpha_k=\frac{1}{K+1}$). Here, we are interested in the weights of distinct graph features and plot them in Figure \ref{weight} (a), where the weight is normalized as $\frac{\sum_{k=0}^K \alpha_k \lambda_t^k}{\sum_{k=0}^K \alpha_k}$ (ratio to the maximum). As increasing the layer $K$, the model depresses the rough features and emphasizes the smoothed features which have been shown important for recommendation in Section 3.1.1. This finding shows the equivalence between exploiting neighborhood signals by stacking layers and emphasizing homophily. We can modify Equation (\ref{graph_filter}) to emphasize heterophily as:
\begin{equation}
\mathbf{O}=\sum_{k=0}^{K}\frac{\mathbf{\mathcal{L}}^k}{K+1}\mathbf{E},
\label{rough_filter}
\end{equation}
where $\mathcal{L}=\mathbf{I}-\mathbf{\hat{A}}$ is the normalized Laplacian matrix that measures the difference between each node and its neighborhood. It is easy to verify that the eigenvalue of $\mathcal{L}$ is $\lambda'_t=1-\lambda_t$, where the rough features are stressed and smoothed features are depressed. It is feasible to combine the two models to capture both the homophily  and heterophily in user behaviour, while we argue that existing work suffers from more limitations that we need to propose a new GCN learning scheme to tackle them:  

\begin{itemize}[leftmargin=10pt]
\item The smoothed features are emphasized by repeating the multiplication of adjacency matrix, which is computationally expensive. 

\item The spectral features are weighted through a polynomial filter in a heuristic way, other designs should be discussed.

\item The features shown useless still remain even stacking 100 layers, indicating the poor ability to denoise graph information.

\item Stacking layers in GCNs results in the over-smoothing issue.

\item The model cannot capture heterophily that might facilitate recommendation as the rough features are heavily depressed.

\end{itemize} 
Evidently, a more flexible and effective design is required to replace the neighborhood-aggregation. Thus, here we plot an ideal design in Figure \ref{weight} (b): the noisy features are unnecessary that should be completely filtered out, while the useful features are reasonably measured according to their importance.

\begin{figure}
\begin{center}
\includegraphics[width=0.48\textwidth]{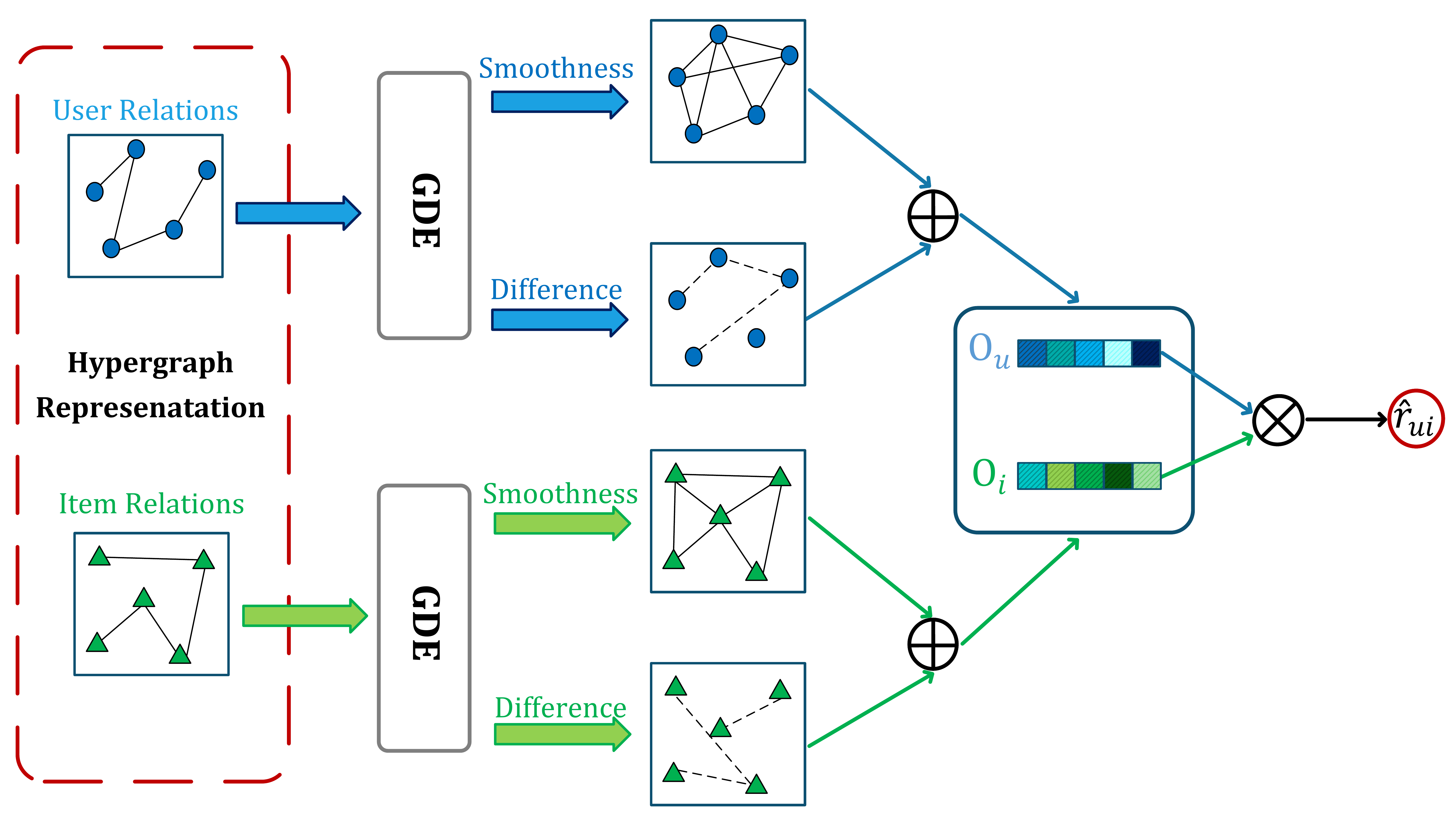}
\vspace{-0.7cm}
\caption{The Framework of our proposed GDE.}
\vspace{-0.3cm}
\label{model}
\end{center}
\end{figure}

\subsection{Proposed Method}
We first formulate the general idea of GDE. Following previous analysis and Equation (\ref{crop}), we can partition an interaction graph $\mathcal{G}$ into smoothed $\mathcal{G}_S$, rough $\mathcal{G}_R$ and noisy $\mathcal{G}_N$ graphs, which are made up of smoothed, rough and noisy spectral features, respectively. The final representations are contributed by the embeddings generated on $\mathcal{G}_S$ and $\mathcal{G}_R$, while the embeddings from $\mathcal{G}_N$ are filtered out. Thus, GDE can be formulated as a band pass filter:
 
\begin{equation}
\gamma(u/i, \lambda_t)\left\{
             \begin{array}{lr}
             \neq0 \quad \quad \, \mathbf{v}_t \in \mathcal{G}_S \, or \, \mathcal{G}_R    \\\\
             
             =0 \quad\quad\ \mathbf{v}_t \in \mathcal{G}_N.\\ 
             \end{array}
\right.
\label{denoiser}
\end{equation}
We replace the polynomial filter by a function $\gamma(\cdot)$. We argue that the importance of a spectral feature not only depends on the variation/eigenvalue but also is related to user/item, thus $\gamma(u, \lambda_t)$ outputs the importance of $t$-th feature to $u$. In practice, we only need to compute the feature from $\mathcal{G}_S$ and $\mathcal{G}_R$ as $\mathcal{G}_N$ is not fed into the model in the first place (\textit{i.e.,} we do not need to consider the situation $\gamma(u, \lambda_t)=0$).\par
To make graph representation more informative and powerful, we represent user-item interactions as hypergraphs:

\begin{equation}
\begin{aligned}
&\mathbf{A}_{U}=\mathbf{D}_u^{\mbox{-}\frac{1}{2}}\mathbf{R}\mathbf{D}_i^{\mbox{-}1}\mathbf{R}^T\mathbf{D}_u^{\mbox{-}\frac{1}{2}} \qquad\in\mathbb{R}^{M\times M},\\
&\mathbf{A}_{I}=\mathbf{D}_i^{\mbox{-}\frac{1}{2}}\mathbf{R}^T\mathbf{D}_u^{\mbox{-}1}\mathbf{R}\mathbf{D}_i^{\mbox{-}\frac{1}{2}} \ \, \qquad\in\mathbb{R}^{N\times N},
\end{aligned}
\label{hyper}
\end{equation}
where $\mathbf{D}_u, \mathbf{D}_i $ are diagonal degree matrices of users and items. Equation (\ref{hyper}) is consistent with the propagation matrix of hypergraph neural network \cite{feng2019hypergraph}, the definition and analysis in the previous section is still applicable here. We treat items (users) as hyper-edges when considering user (item) relations.

\subsubsection{Graph Denoising Encoder (GDE)}
We illustrate the proposed GDE in Figure \ref{model}, where we propagate embeddings on hypergraphs which emphasize the neighborhood smoothness and difference. The embeddings generated on the smoothed hypergraphs are formulated as follows: 
\begin{equation}
\begin{aligned}
&\mathbf{H}_{U}^{(s)}=\left( \mathbf{P}^{(s)} \odot\gamma\left(\mathcal{U}, \pi^{(s)}\right){\mathbf{P}^{(s)}}^T\right) \mathbf{E}_U,\\
&\mathbf{H}_{I}^{(s)}=\left(\mathbf{Q}^{(s)}\odot\gamma\left(\mathcal{I}, \sigma^{(s)}\right){\mathbf{Q}^{(s)}}^T\right)\mathbf{E}_I,
\end{aligned}
\label{smooth_capture}
\end{equation}
where $\{\mathbf{P}^{(s)}\in\mathbb{R}^{M\times m_1}$, $\pi^{(s)}\in\mathbb{R}^{m_1}\}$, $\{\mathbf{Q}^{(s)}\in\mathbb{R}^{N\times n_1,}$, $\sigma^{(s)}\in\mathbb{R}^{n_1}\}$ are top $m_1$ and $n_1$ smoothed \{features,  eigenvalues\} for user and item relations ($\mathbf{A}_U$ and $\mathbf{A}_I$), respectively. The term in parentheses represents the node relations on the smoothed graph; $\gamma(\cdot)$ outputs the importance of distinct features to users/items, $\odot$ stands for the element-wise multiplication. $\mathbf{E}_U$ and $\mathbf{E}_I$ are embedding matrices for users and items, respectively. To improve generalization on test sets, we randomly drop out the node relations with a ratio $p\in[0,1]$. Similarly, we propagate embeddings on the rough hypergraphs with top $m_2+n_2$ rough features to learn the heterophily:
\begin{equation}
\begin{aligned}
&\mathbf{H}_{U}^{(r)}=\left(\mathbf{P}^{(r)} \odot\gamma\left(\mathcal{U}, \pi^{(r)}\right){\mathbf{P}^{(r)}}^T\right) \mathbf{E}_U,\\
&\mathbf{H}_{I}^{(r)}=\left(\mathbf{Q}^{(r)}\odot\gamma\left(\mathcal{I}, \sigma^{(r)}\right){\mathbf{Q}^{(s)}}^T\right) \mathbf{E}_I.
\end{aligned}
\label{rough_capture}
\end{equation}  
Similarly, $\{\mathbf{P}^{(r)}$, $\pi^{(r)}$\}, $\{\mathbf{Q}^{(r)}$, $\sigma^{(r)}\}$ are top $m_2$ and $n_2$ rough \{features,  eigenvalues\} for user and item relations, respectively. We let $m=m_1+m_2$, $n=n_1+n_2$, and generate the final embeddings by:

\begin{equation}
\begin{aligned}
&\mathbf{O}_{U}=\mathbf{pooling}\left(\mathbf{H}^{(s)}_U,\mathbf{H}^{(r)}_U\right), \\
&\mathbf{O}_{I}=\mathbf{pooling}\left(\mathbf{H}^{(s)}_I,\mathbf{H}^{(r)}_I\right).
\end{aligned}
\label{final_spatial}
\end{equation}
To keep the model simple and avoid bringing additional complexity, we take summation for pooling function. According to our analysis in Section 3.1.2, stacking layers in GCNs is essentially reweighting the spectral features. Since a single-layer GDE is capable of reasonably weighting the important graph features for recommendation, it is unnecessary to stack more layers.   

\subsubsection{Measuring the Importance of Graph Features}
There are two directions for the design of $\gamma(\cdot)$: a dynamic design by parameterizing $\gamma(\cdot)$, or a well motivated static design by manually adjusting hyper-parameters without introducing parameters. Here, we introduce a instantiation for each of them. For simplicity, we define $\mathbf{P}=[\mathbf{P}^{(s)} \left| \right| \mathbf{P}^{(r)}]$, $\pi=[\pi^{(s)} \left| \right| \pi^{(r)}]$ and $\mathbf{Q}=[\mathbf{Q}^{(s)} \left| \right|\mathbf{Q}^{(r)}]$, $\sigma=[\sigma^{(s)} \left| \right| \sigma^{(r)}]$ ($\left |\right|$ is the concatenate operation).\\   
$\bullet$ \textbf{Option \uppercase\expandafter{\romannumeral1}}. We first encode variation into spectral features and define variation-encoded feature matrices as: $\mathbf{P}'=\mathbf{P}diag(\pi)$, $\mathbf{Q}'=\mathbf{Q}diag(\sigma)$. The importance of a spectral feature to a user is generated via attention mechanism:
\begin{equation}
\gamma(u, \pi_g)=\sigma\left(\mathbf{a}^T\left[\mathbf{W}_U^{(1)}\mathbf{P}'^T_u \left | \right | \mathbf{W}_U^{(2)}\mathbf{P}'_g \right] \right),
\label{attention_weight}
\end{equation}
where $\mathbf{P}'^T_u$ is a feature vector of $u$, $\mathbf{P}'_g$ is $g$-th spectral feature; $\mathbf{W}_U^{(1)}\in \mathbb{R}^{d_1 \times m}$, $\mathbf{W}_U^{(2)}\in \mathbb{R}^{d_1 \times M}$ are transform matrices, attention mechanism is parameterized as a single-layer neural network where $\mathbf{a}\in\mathbb{R}^{2d_1}$. We can learn the importance of a spectral feature to an item similarly. Finally, we normalize the scores across features using the softmax function. \\
$\bullet$ \textbf{Option \uppercase\expandafter{\romannumeral2}}. We first analyze what static design could lead to better results. By considering $\gamma(\cdot)$ as a one variable continuous function of eigenvalues (\textit{i.e.,} ignore the effect from user/item first), we can rewrite the term in parentheses in Equations (\ref{smooth_capture}) and (\ref{rough_capture}) as follows according to the Taylor series:
\begin{equation}
\mathbf{P} diag(\sum_{k=0}^K \alpha_k \pi^k)\mathbf{P}^T=\sum_{k=0}^K \alpha_k \mathbf{P} diag(\pi^k)\mathbf{P}^T=\sum_{k=0}^K \alpha_k {\mathbf{\bar{A}}_U^{k}},
\label{rewrite}
\end{equation}
where ${\mathbf{\bar{A}}_U^{k}}$ can be considered as an adjacency matrix with noisy features being removed, $\mathbf{\bar{A}}_U^{k}=\mathbf{A}_U^{k}$ when $m\rightarrow M$. $K$ is the highest order with non-zero derivative, $\alpha_k=\frac{\gamma^{(k)}(0)}{k!}$ is the coefficient of $k$-th order Maclaurin expansion. On the other hand, from a spatial perspective, $K$ is also the order of the farthest incorporated neighborhood and $\alpha_k$ is the contribution of $k$-th order neighborhood. Intuitively, we hope the model can capture neighbor signals as far as possible with positive contributions to user/item representations, implying that $\alpha_k>0$, $K\rightarrow\infty$. In other words, $\gamma(\cdot)$ should be infinitely differentiable whose any-order derivative is positive. To satisfy this condition and after extensive experiments (shown in Section 4.4.2), we use a exponential kernel: $\gamma(\pi)=e^{\beta\pi}$, where $\beta\in\mathbb{R}$ controls the extent of the emphasis over different features (\textit{i.e.,} a larger (smaller) $\beta$ emphasize the neighbor smoothing (difference) more). We can rewrite Equation (\ref{rewrite}) as $\mathbf{A}_U'=\sum_{k=0}^{\infty} \frac{\beta^k}{k!} {\mathbf{\bar{A}}_U^{k}}$, and is comparable to a GCN with infinite layers if we further rephrase GDE in the form of the GCN paradigm for CF: 
\begin{equation}
\begin{aligned}
&\mathbf{H}^{(k+1)}=\mathbf{\bar{A}}_U \mathbf{H}^{(k)},\\
&\mathbf{O}_U=\lim_{K\rightarrow \infty} \sum_{k=0}^K \frac{\beta^k}{k!} \mathbf{H}^{(k)}.
\end{aligned}
\end{equation}
Furthermore, the importance of a spectral feature might vary on different users/items as well. For instance, the users/items with low degrees are isolated on the graph, thus the smoothing effect should be emphasized more than the nodes with high degrees. To this end, we modify as $\gamma(u, \pi_g)=e^{(\beta+({\rm -} log (d_u)))\pi_g}$ to adapt to different users/items. \par

In our experiments, we choose \textbf{Option \uppercase\expandafter{\romannumeral2}} as it shows improvement over \textbf{Option \uppercase\expandafter{\romannumeral1}} across all datasets. Here, we attempt to analyze the limitations of an adaptive design. Firstly, with a parameterized design, the terms in parentheses in Equations (\ref{smooth_capture}) and (\ref{rough_capture}) need to be repeated during each epoch of training, which is computationally expensive and unnecessary when using an static design. Secondly, we notice that introducing parameters to model the importance of features results in even worse convergence and accuracy. We speculate the reason is due to the sparseness of the datasets. Unlike other tasks, the available data for CF is only the user/item ID, which is difficult to learn the data intrinsic characteristics in an adaptive manner. We will compare the two designs in Section 4.4.2.

\subsection{Discussion}

\subsubsection{Over-Smoothing}   
\begin{definition}
A model suffers from the over-smoothing if any spectral features dominate as the model layer $K$ is large enough:
\begin{equation}
\lim_{K\rightarrow\infty}\frac{\left|\gamma(\lambda_t)\right|}{\mathbf{max}\{\left|\gamma(\lambda_1)\right|,\cdots, \left|\gamma(\lambda_{M+N})\right|\}}\rightarrow0.
\end{equation} 
\end{definition}
Over-smoothing in GCNs \cite{li2018deeper,liu2020towards} refers to overweight of the smoothest feature when increasing the layer $K$, eventually all features except the smoothest one loses and it results in the same user/item representations. Most GCN-based CF methods suffer from this limitation and remain shallow, take LightGCN as an example: 
\begin{equation}
\begin{aligned}
&{\rm LightGCN}:\,\lim_{\substack{\alpha_k=\frac{1}{K+1}\\ K\rightarrow\infty}}\frac{\gamma(\lambda_t)}{\gamma(\lambda_{\rm{max}})}=\frac{\sum_{k=0}^{K}\frac{\lambda_t^k}{K+1}}{\sum_{k=0}^{K}\frac{1}{K+1}}\rightarrow0.
\end{aligned}
\end{equation}
The key to prevent over-smoothing is to properly and reasonably model the importance of features too assure any features are not overweighted. On the other hand, instead of controlling the weight through neighborhood aggregation, we adjust the weight of different features through a flexible and light design (\textit{i.e.,} \textbf{Option \uppercase\expandafter{\romannumeral2}}). According to Definition 2, it is easy to verify that GDE does not suffer from over-smoothing: $\frac{e^{\beta \pi_g}}{e^{\beta \pi_{\rm max}}}\neq0$, $\frac{e^{\beta \sigma_h}}{e^{\beta \sigma_{\rm max}}}\neq0$.

\subsubsection{Time Complexity}    
The complexity of our model mainly comes from retrieving of required graph features (preprocessing) and the training. We can calculate spectral features through algorithms such as Lanczos method and LOBPCG \cite{paratte2016fast,stathopoulos2002block} with GPU implementation. For instance, the complexity of Lanczos method is $\mathcal{O}(m^2M+m|\mathcal{E}_{\mathbf{A}_{U}}|+n^2N+n|\mathcal{E}_{\mathbf{A}_{I}}|)$ \cite{paratte2016fast}, where $\mathbf{A}_U$ and $\mathbf{A}_I$ are sparse, and $m\ll M$, $n\ll N$, $\mathcal{E}_{\mathbf{A}_{U}}$ and $\mathcal{E}_{\mathbf{A}_{I}}$ are edges of $\mathbf{A}_U$ and $\mathbf{A}_I$, respectively. Since $\mathbf{A}_U$ and $\mathbf{A}_I$ are sparse and required features only account for a small portion, the complexity can be controlled at a low level. During training, each user/item can be considered as a multi-hot vector, and the final embedding is generated in a node level rather than a matrix level. The complexity for GDE is $\mathcal{O}((1-p)(M+N)dc\left|\mathbf{R}^+\right|)$, where $c$ is the number of epochs. 

\subsection{Optimization}
\begin{figure} \centering 
\subfigure[The gradients on negative samples vanishes more quickly on LightGCN than MF.] {  
\includegraphics[width=0.46\columnwidth]{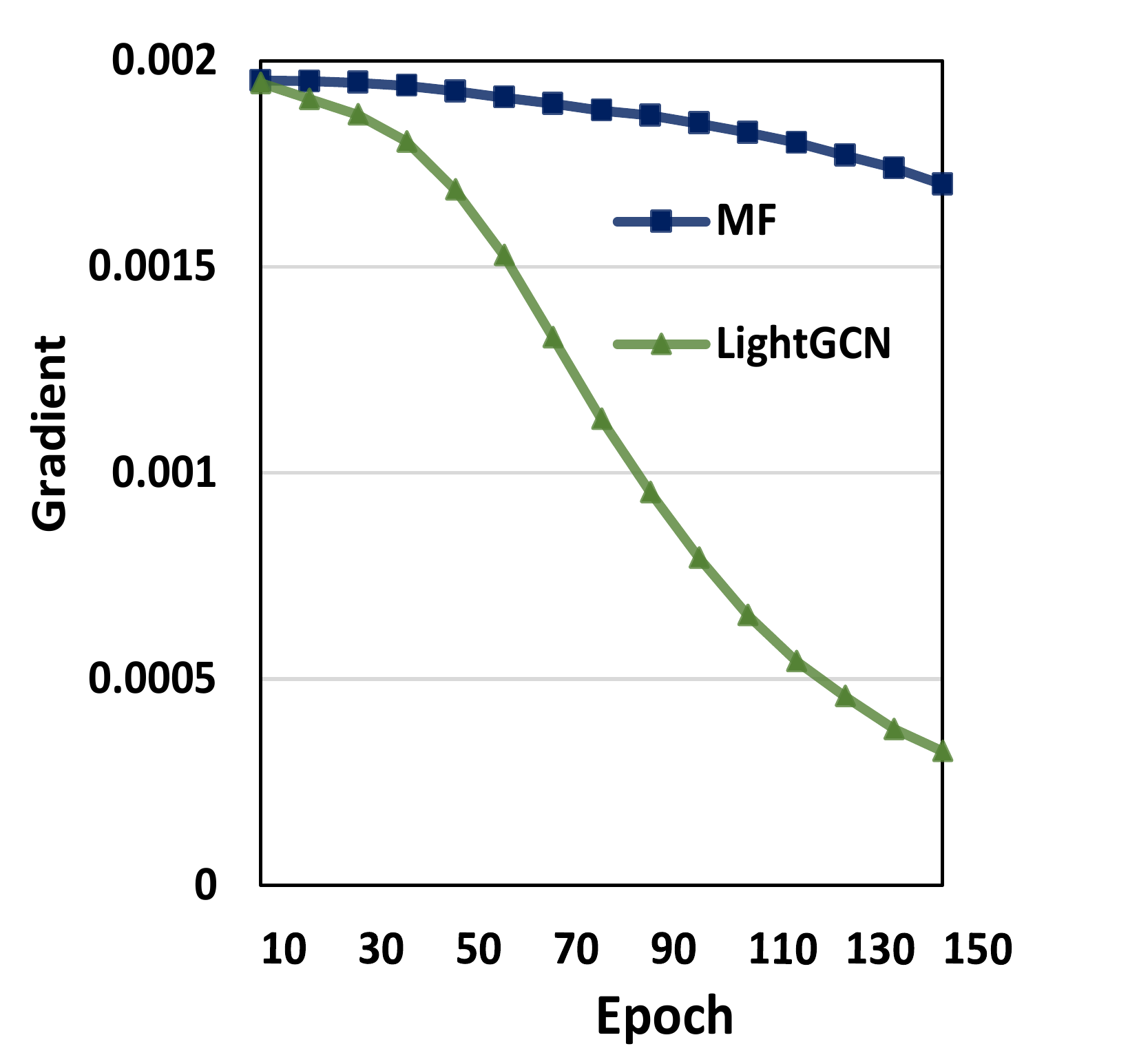} 
}  
\hspace{0cm}
\subfigure[This issue is eased by adaptively adjusting the gradients on negative samples. ] {  
\includegraphics[width=0.46\columnwidth]{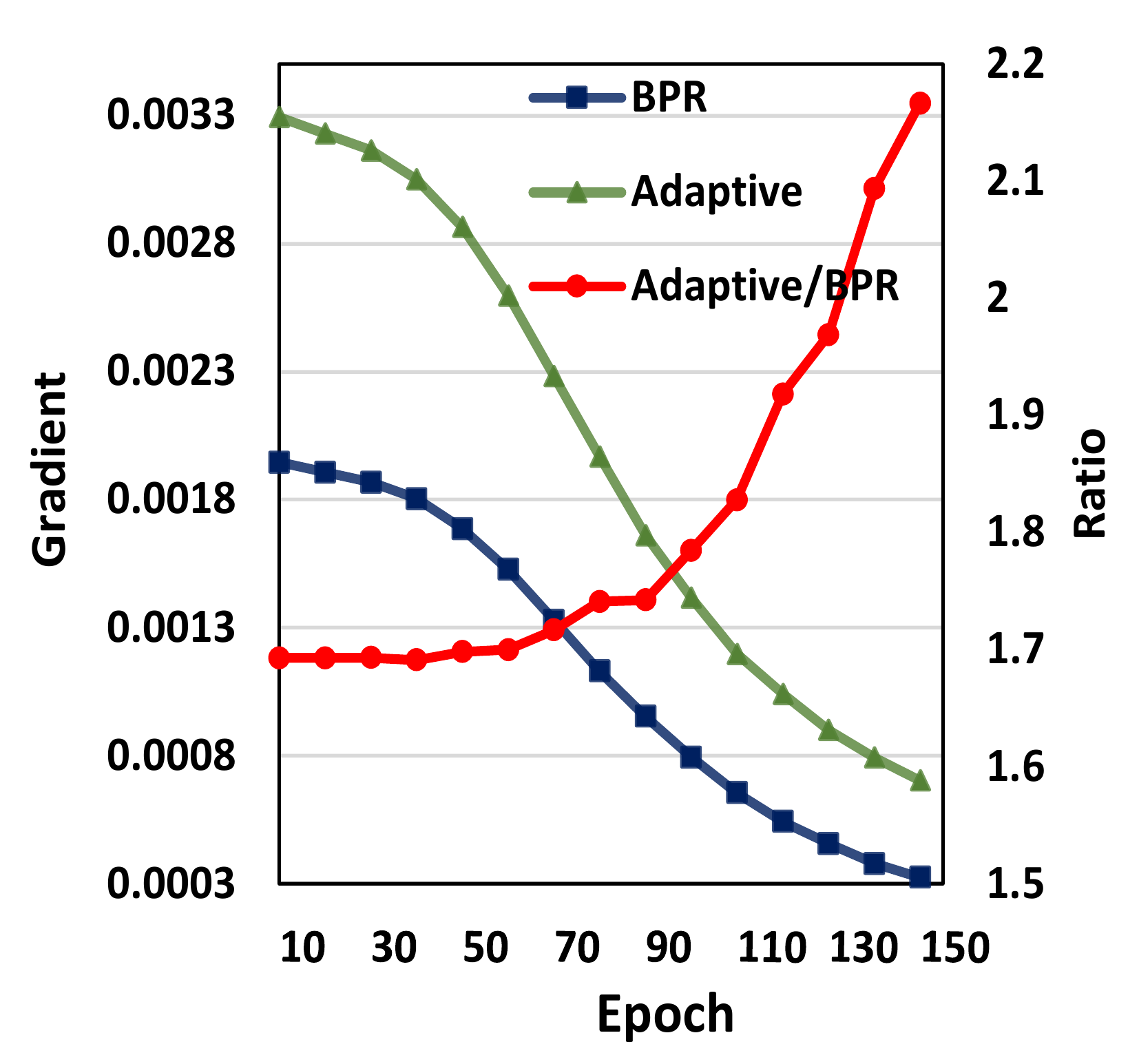} 
}
\vspace{-0.2cm}
\caption{A training issue of pairwise learning and a solution to it.}  
\label{adaptive}
\vspace{-0.2cm}
\end{figure}

It has been reported that some GCN-based methods show slow training convergence. We argue that the issue lies in the commonly used BPR loss for pairwise learning \cite{rendle2009bpr}:
\begin{equation}
\mathcal{L}_{BPR}=-\sum_{(u,i,j)\in{T}}\ln\sigma(\mathbf{\hat{r}}_{ui}-\mathbf{\hat{r}}_{uj}),
\end{equation}
where $T=\{(u,i,j)|(u,i)\in{\mathbf{R}^+}, (u,j)\in{\mathbf{R}^{\mbox{-}}}\}$, $\sigma{(\cdot)}$ is a sigmoid function. The loss optimizes based on sampled triples by maximizing the difference between observed interactions and unobserved interactions. Rendle and Frendenthaler \cite{rendle2014improving} show that the positive rating $\mathbf{\hat{r}}_{ui}$ increases quickly due to the tailed item distributions (\textit{i.e.,} observed interactions are more likely to be sampled than unobserved ones). As a result, the term controlling the gradient magnitude $\left(1-\sigma(\mathbf{\hat{r}}_{ui}-\mathbf{\hat{r}}_{uj})\right)$ decreases quickly and prevents the model from learning from (negative) training pairs. Figure \ref{adaptive} (a) shows that LightGCN suffers more from this issue than MF, which explains why it requires many epochs to converge. To avoid bringing additional complexity, we propose to adaptively adjust the gradient to expedite model training:
\begin{equation}
\delta_{uj}=1-\log(1-{\rm min}(\sigma(\mathbf{\hat{r}}_{uj}),\xi)).
\end{equation} 
Since this issue is due to the over-sampling of positive items, we adjust the gradients over negative samples. $\delta_{uj}$ is a self-paced but non-trainable parameter changing according to $\mathbf{\hat{r}}_{uj}$; when $\sigma(\mathbf{\hat{r}}_{uj})$ deviates too far from 0, $\delta_{uj}$ becomes large to accelerate the model training. ${\rm min}(\sigma(\hat{r}_{uj}),\xi)$ is to prevent $\delta_{uj}$ being too large, and we set $\xi=0.99$. The loss function is enhanced as follows:
\begin{equation}
\mathcal{L}_{adapt}=-\sum_{(u,i,j)\in{T}}\ln\sigma(\mathbf{\hat{r}}_{ui}-\delta_{uj}\mathbf{\hat{r}}_{uj})+\lambda\left \| \Theta \right \| ^2,
\label{new_loss}
\end{equation} 
where $\lambda$ controls regularization strength. We compare two loss functions in Figure \ref{adaptive} (b) by showing the gradient with respect to $\mathbf{\hat{r}}_{uj}$. Obviously, the negative items consistently receive larger gradients from the adaptive loss than the plain BPR loss, thereby helping accelerate training. 

\section{Experiments}
In this section, we comprehensively evaluate GDE. In particular, we aim to answer the following research questions:
\begin{itemize}[leftmargin=10pt]
\item \textbf{RQ1}: Does GDE outperform other baselines?
\item \textbf{RQ2}: Is GDE more efficient than GCN-based methods?
\item \textbf{RQ3}: Does the proposed designs show positive effects? How do hyper-parameters affect the performance? 
\end{itemize}

\subsection{Experimental Setup}
\subsubsection{Datasets} The descriptions of datasets are listed as follows. The statistics of all five datasets are summarized in Table \ref{datasets}.

\begin{itemize}[leftmargin=10pt]
\item \textbf{Pinterest}: This is an implicit feedback dataset \cite{he2017neural} for content-based image recommendation, where users can pin image they are interested in.
\item \textbf{CiteULike-a}: This dataset\footnote{https://github.com/js05212/citeulike-a} is collected from CiteULike which allows users to create their own collections of articles.
\item \textbf{MovieLens}: These two datasets (1M and 100K)\footnote{https://grouplens.org/datasets/movielens/} have been widely used to evaluate CF algorithms. Since it is an explicit feedback dataset while we focus on implicit feedback, we hide all ratings.
 \item \textbf{Gowalla}: The interactions in this dataset \cite{wang2019neural} are check-ins which record the locations the user has visited.
\end{itemize}

\begin{table}
\centering
\caption{Statistics of datasets}
\vspace{-0.2cm}
\begin{tabular}{lcccc}
\toprule
Datasets&\#User&\#Item &\#Interactions &Density\%\\
\midrule
CiteULike-a&5,551&16,981&210,537&0.223\\
MovieLens-1M&6,040&3,952&1,000,209&4.190\\
MovieLens-100K&943&1,682&100,000&6.305\\
Pinterest&37,501&9,836&1,025,709&0.278\\
Gowalla&29,858&40,981&1,027,370&0.084\\
\bottomrule
\label{datasets}
\vspace{-0.6cm}
\end{tabular}
\end{table}

\begin{table*}[]
\caption{Overall performance comparison. Improv.\% denotes the improvements over the best baselines. }
\vspace{-0.2cm}
\scalebox{0.83}{
\begin{tabular}{c|cc|cc|cc|cc|cc}
\hline
              & \multicolumn{2}{c|}{\textbf{CiteULike}} & \multicolumn{2}{c|}{\textbf{Pinterest}} & \multicolumn{2}{c|}{\textbf{MovieLens-1M}} & \multicolumn{2}{c|}{\textbf{Gowalla}} & \multicolumn{2}{c}{\textbf{MovieLens-100K}} \\ \cline{2-11} 
              & \textbf{nDCG@20}  & \textbf{Recall@20}  & \textbf{nDCG@20}  & \textbf{Recall@20}  & \textbf{nDCG@20}    & \textbf{Recall@20}   & \textbf{nDCG@20} & \textbf{Recall@20} & \textbf{nDCG@20}    & \textbf{Recall@20}    \\ \hline
BPR           & 0.0591            & 0.0527              & 0.0861            & 0.0809              & 0.4849              & 0.4578               & 0.0907           & 0.0743             & 0.4935              & 0.4641                \\
Ease          & 0.0846            & 0.0801              & 0.0695            & 0.0639              & 0.3249              & 0.3000               & 0.0670           & 0.0642             & 0.3523              & 0.3214                \\ \hline
LCFN          & 0.0662            & 0.0590              & 0.0937            & 0.0873              & 0.5197              & 0.4898               & 0.1132           & 0.0980             & 0.5199              & 0.4898                \\
GF-CF          & 0.0836            & 0.0811              & 0.0776            & 0.0755              & 0.4789              & 0.4562               & 0.0537           & 0.0567             & 0.4048              & 0.3793                \\
ELGN          & 0.1125      &  0.1027        &  0.1176      & \underline{ 0.1098}        & \underline{ 0.5418}        & \underline{ 0.5133}         & 0.1249           & 0.1138             & 0.5347              & 0.5100                \\
LightGCN      & \underline{0.1149}            & \underline{0.1066}              & 0.1143            & 0.1069              & 0.5261              & 0.5031               & 0.1327           & 0.1224 & \underline{ 0.5418}        & \underline{ 0.5133}          \\
SGL-ED        & 0.1070            & 0.0985              & \underline{0.1185}            & 0.1094 & 0.5314              & 0.5035               & \underline{ 0.1561}     & \underline{ 0.1353}       & 0.5321              & 0.5044                \\ \hline
GDE           & \textbf{0.1339*}   & \textbf{0.1224*}     & \textbf{0.1240*}   & \textbf{0.1147*}     & \textbf{0.5715*}     & \textbf{0.5423*}      & \textbf{0.1632*}  & \textbf{0.1449*}    & \textbf{0.5731*}     & \textbf{0.5400*}       \\
GDE-d         & 0.1126            & 0.1026              & 0.1157            & 0.1080              & 0.5578              & 0.5306               & 0.1462           & 0.1341             & 0.5582 & 0.5280                \\
Improv.\% & +16.54            & +14.82              & +3.29             & +4.46               & +5.48               & +5.65                & +4.55            & +7.10             & +5.77 & +5.20                 \\ 
$p${\rm -}value & 4.10e{\rm -}9           & 2.77e{\rm -}8             & 2.94e{\rm -}4            & 3.71e{\rm -}7              & 6.71e{\rm -}7               & 5.44e{\rm -}4                & 6.15e{\rm -}5        & 5.44e{\rm -}4              & 3.88e{\rm -}6  & 4.29e{\rm -}6                  \\ \hline
\end{tabular}}
\label{table_comparison}
\end{table*}

\subsubsection{Evaluation Metrics} We adopt two widely-used metrics: Recall and nDCG for personalized ranking \cite{jarvelin2002cumulated}. Recall measures the ratio of recommended items in the test set; nDCG considers the position of items by assigning a higher weight to the item ranking higher. The recommendation list is generated by ranking unobserved items and truncating at position $k$. As the success of GCNs lies in the ability of exploiting high-order neighbor to tackle data sparsity which is common in practice, we use only 20\% of the user-item pairs for training to evaluate the model stability with limited interactions, and leave the remaining for test; we randomly select 5\% from the training data as validation set for hyper-parameter tuning. We report the average accuracy on test sets. 

\subsubsection{Baselines} We compare our proposed method with the following CF methods. The architecture settings are based on the reported results in each paper:
\begin{itemize}[leftmargin=10pt]
\item BPR \cite{rendle2009bpr}: This method proposes a pair-wise ranking loss by maximizing the difference between observed and unobserved interactions. 
\item EASE \cite{steck2019embarrassingly}: This is a neighborhood-based method which is considered as a SLIM \cite{ning2011slim} variant with a closed form solution. 
\item LCFN \cite{yu2020graph}: This model proposes a low pass graph convolution to replace the vanilla graph convolution and initializes the embeddings with pretrained MF. We set $F=0.1$ on CiteULike and $F=0.005$ on other datasets.  
\item LightGCN \cite{he2020lightgcn}: Unlike other GCN methods, this model exploits a light GCN model for CF by removing non-linear activation functions and transformation layers where the model complexity is the same as MF.
\item GF-CF \cite{shen2021powerful}: This is a GCN-based CF method without model optimization, exploiting low frequency components and has a low time complexity. 
\item EGLN \cite{yang2021enhanced}: This model uses an adaptive user-item graph structure and deigns a local-global consistency optimization function via mutual information maximization to better serve CF. We set $\alpha=0.1$ and $\beta=0.1$.
\item SGL-ED \cite{wu2020self}: This model contrasts different node views that are generated by randomly masking the edge connections on the graph, and incorporate the proposed self-supervised loss into LightGCN \cite{he2020lightgcn}. We set $\tau=0.2$, $\lambda_1=0.1$ and $p=0.1$. 
\end{itemize}
We remove popular GCN-based methods such as GCMC \cite{berg2017graph}, SpectralCF \cite{zheng2018spectral}, Pinsage \cite{ying2018graph}, NGCF \cite{wang2019neural} as the baselines above have shown superiority over them.

\subsubsection{Implementation details} We implemented the proposed model based on PyTorch$\footnote{https://pytorch.org/}$, and released the code on Github$\footnote{https://github.com/tanatosuu/GDE}$. For all models, the optimizer is SGD; the embedding size $d$ is set to 64 and $d_1=16$; the regularization rate is set to 0.01 on all datasets; the learning rate is tuned amongst $\{0.001,0.005,0.01,\cdots \}$; the drop ratio of node relations is tune amongst $p=\{0.1, 0.2, \cdots\}$; without specification, the model parameters are initialized with Xavier Initialization \cite{glorot2010understanding}; the batch size is set to 256. We report the hyper-paramter setting: $\frac{m}{M}/\frac{n}{N}=\{0.01, 0.05, 0.1, 0.2, \cdots\}$, $\beta=\{1, 1.5, 2, 2.5, \cdots \}$ in the next subsection.

\subsection{Overall Comparison (RQ1)}
We report the performance of baselines and our GDE variants in Table \ref{table_comparison}, where GDE-d is a GDE variant using Option \uppercase\expandafter{\romannumeral1}. We observe the followings: 

\begin{itemize}[leftmargin=10pt]
\item Overall, GCN-based methods outperform traditional CF methods when data suffers from extreme sparsity, indicating the effectiveness of GCNs to tackle the data sparsity. EGLN, LightGCN, and SGL-ED alternately achieve the best baseline, while our proposed GDE consistently outperforms all baselines across all datasets, indicating the superiority and stability of our proposed method.    

\item Among GCN-based methods, LCFN and GF-CF show relatively poor performance compared with other methods. We speculate that the parameterized kernel used in LCFN fails to learn the importance of low frequency components, such an design even reduces the accuracy and hinders the convergence. As for GF-CF, a reasonable explanation is GF-CF fails to perform stably with limited interactions due to the lack of model optimization. The sparse setting adopted in our work increases the difficulty to learn from data and to perform stably, which is also critical to evaluate recommendation models.    

\item Since our work improves the GCN architecture for CF, it is more fair to compare with pure GCN-based methods such as LightGCN. The improvement of GDE over LighGCN is more significant on sparse data (\textit{e.g.,} 23.0\% on Gowalla in terms of nDCG@20) than dense data (\textit{e.g.,} 5.8\% on 100K), indicating the effectiveness of GDE to tackle data sparsity.

\item Although GDE-d shows competitive performance over baselines, it still significantly underperforms GDE, indicating that it is difficult to learn the data characteristic in an adaptive manner, while a well motived static design might result in a promising accuracy. We will compare the two designs in terms of complexity and performance in detail in the latter section.

\end{itemize}

\begin{table}[]
\caption{The comparison of the training time (seconds) per epoch on Gowalla with batch-size=256.}
\vspace{-0.2cm}
\scalebox{0.89}{
\begin{tabular}{c|c|c|c|c|c}
\hline
         & CiteULike & Pinterest & ML-1M & Gowalla & Epochs \\ \hline
ELGN     & 13.90      & 268.7      & 26.5       & 770.1   & 260   \\
LightGCN & 1.17      & 7.7      & 8.0       & 9.8  &600 \\
LCFN     & 33.71     & 154.3     & 4.8       & 269.7  &350  \\
BPR      & 0.44      & 1.9       & 1.8       & 3.3  &450     \\\hline
GDE    & 0.39      & 1.8       & 1.8       & 3.1   &120    \\ 
GDE-d    & 3.15      & 26.7       & 3.3       & 34.5  &180    \\ \hline
\end{tabular}}
\label{time_comparison}
\vspace{-0.2cm}
\end{table}

\begin{figure*} \centering 
\subfigure[CiteULike] {  
\includegraphics[width=0.49\columnwidth]{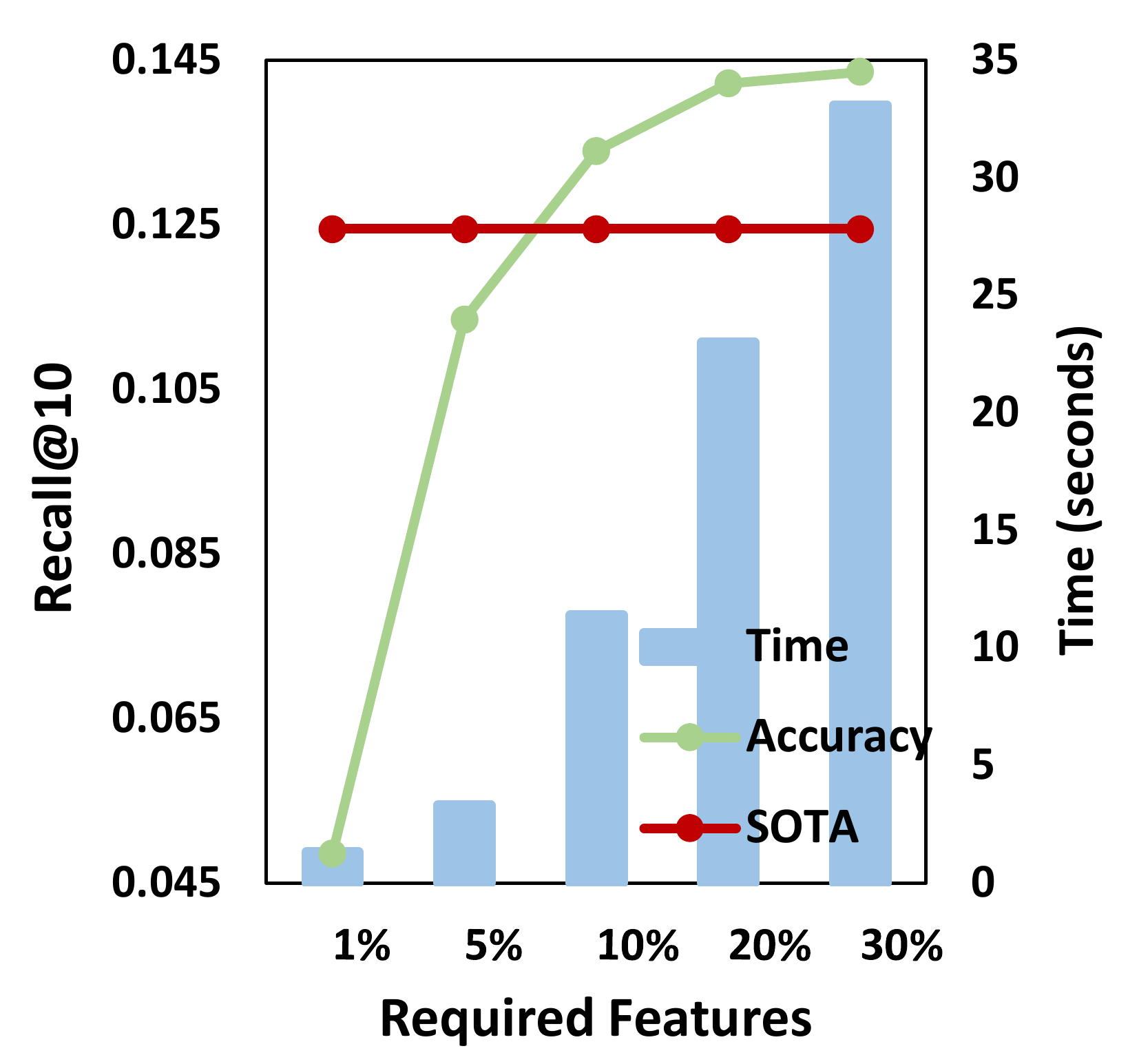} 
}  
\hspace{0.02cm}
\subfigure[Pinterest ] {  
\includegraphics[width=0.49\columnwidth]{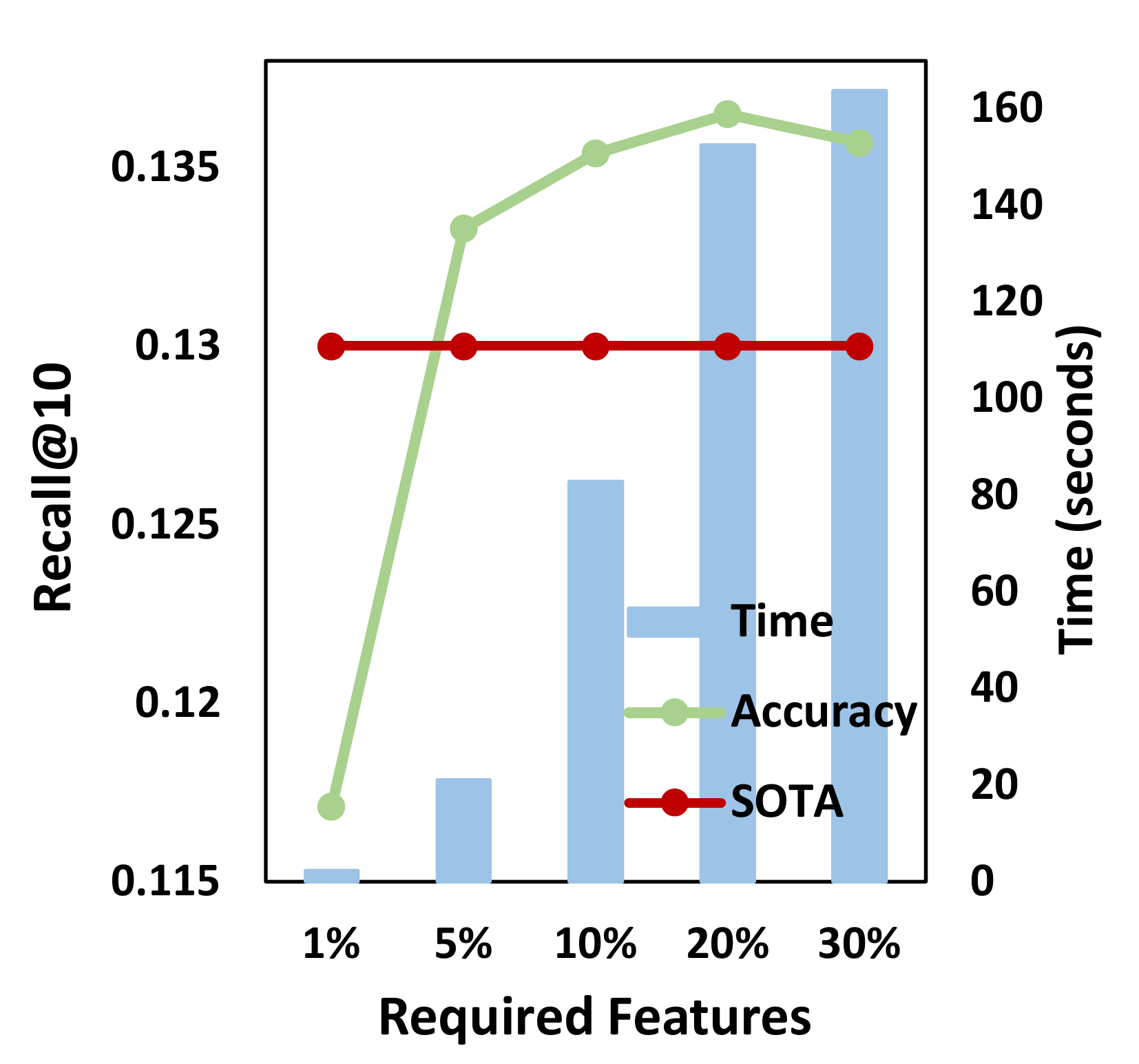} 
}
\hspace{0.02cm}
\subfigure[MovieLens-1M ] {  
\includegraphics[width=0.49\columnwidth]{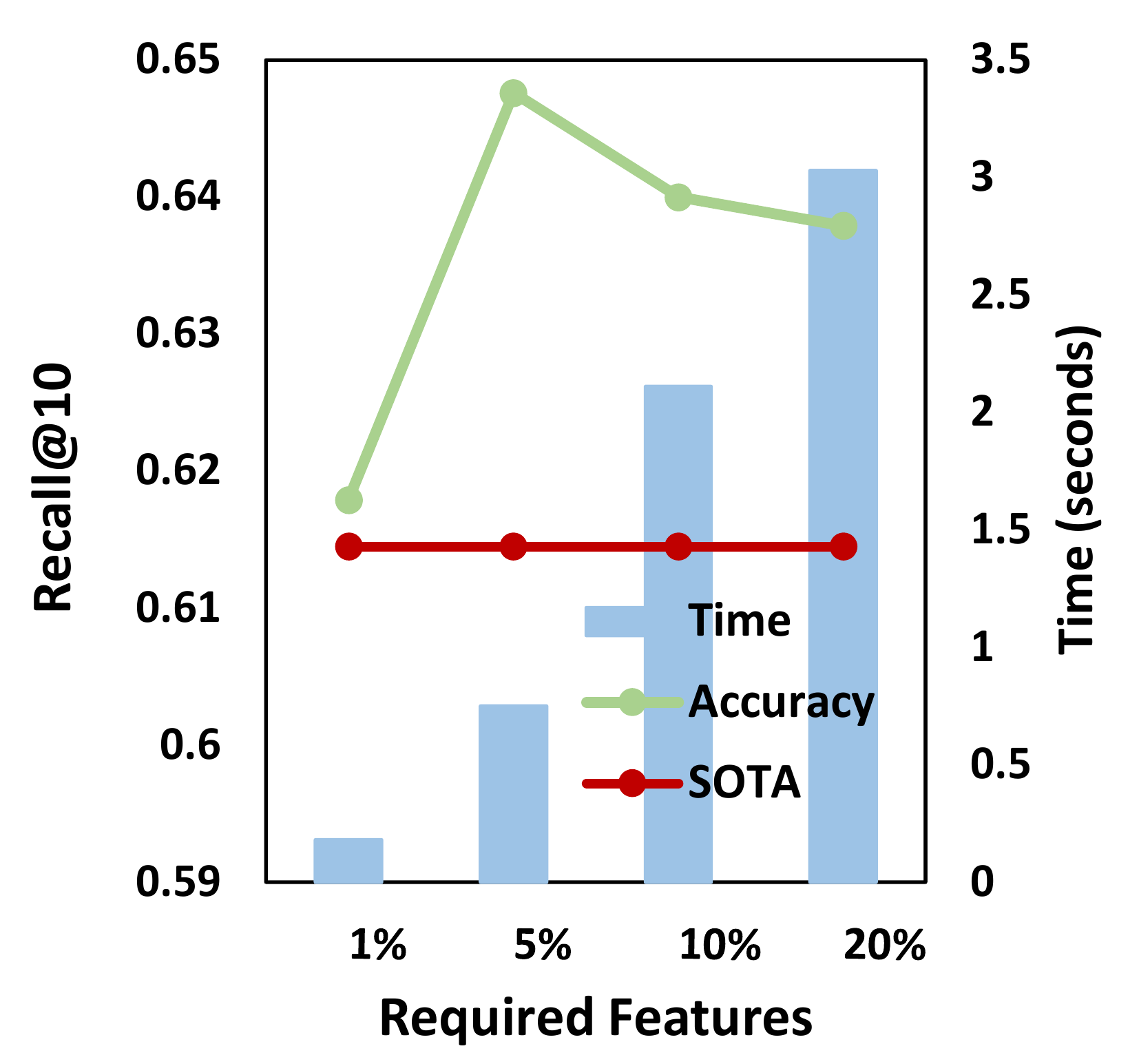} 
}
\hspace{0.02cm}
\subfigure[Gowalla ] {  
\includegraphics[width=0.49\columnwidth]{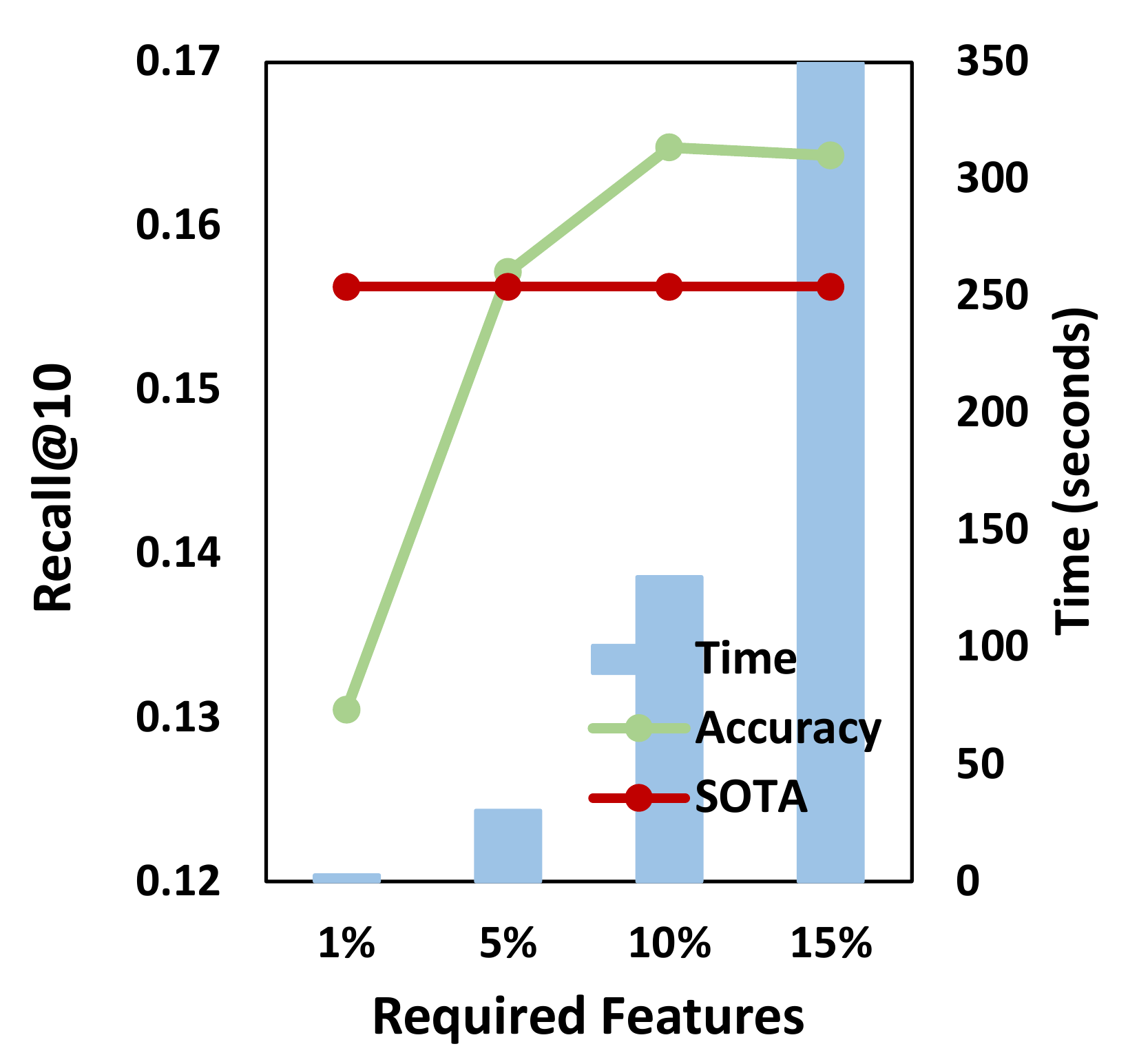} 
}
\vspace{-0.4cm}
\caption{How the accuracy and preprocessing time (y-axis) change with the required spectral features (x-axis). SOTA represents the accuracy of the best baseline (\textit{i.e.,} LightGCN on CiteULike, SGL-ED on Pinterest and Gowalla, ELGN on ML-1M).}  
\vspace{-0.4cm}
\label{effect_t}
\end{figure*}

\begin{figure} \centering 
\vspace{-0.2cm}
\subfigure[Accuracy] {  
\includegraphics[width=0.46\columnwidth]{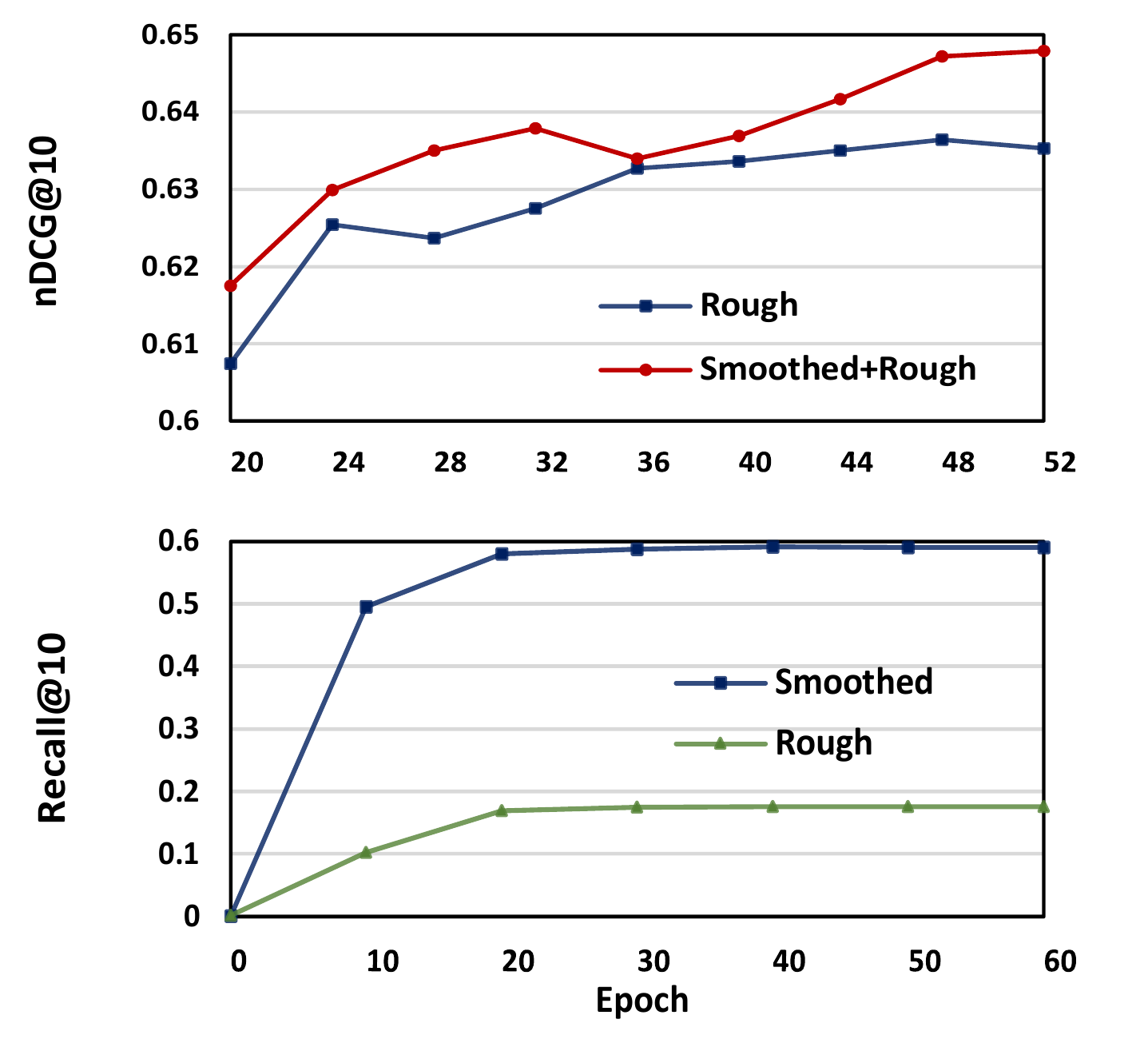} 
}  
\hspace{-0cm}
\subfigure[Training Loss ] {  
\includegraphics[width=0.46\columnwidth]{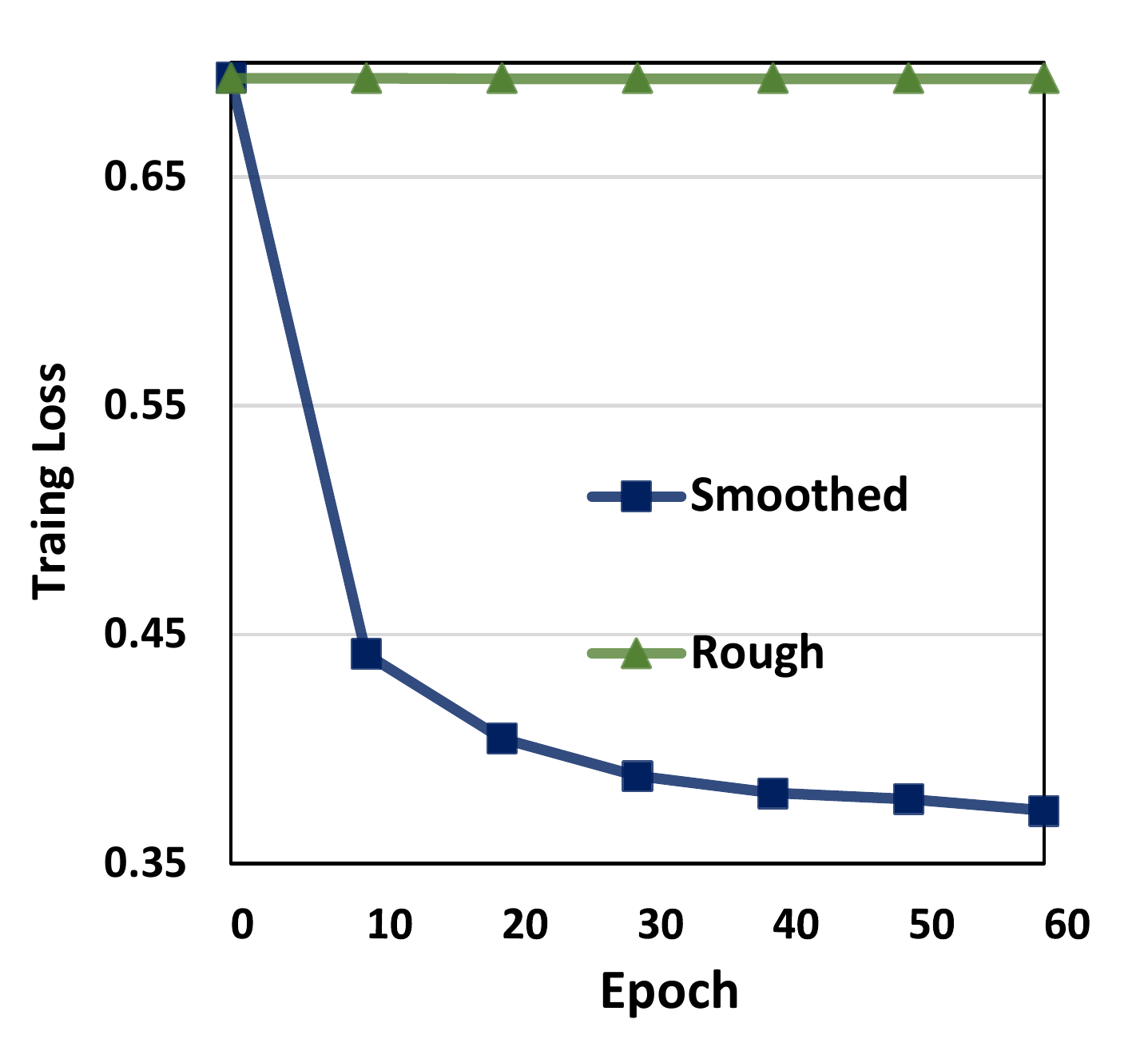} 
}
\vspace{-0.2cm}
\caption{How rough and smoothed features affect accuracy and training loss as the training proceeds.}  
\label{rough_smooth}
\vspace{-0.2cm}
\end{figure}

\subsection{Efficiency of GDE (RQ2)}

Figure \ref{effect_t} shows the preprocessing time. We can see the model shows superior performance with only a small portion of spectral features. For instance, GDE outperforms the best baseline with less than 10\%, 5\%, 1\%, 5\% spectral features on the datasets in Figure \ref{effect_t} (a) to (d), respectively, that justifies our previous analysis that only a very small portion of graph features is useful for recommendation. In practice, we can adjust the number of graph features in order to balance between the computational complexity and model accuracy. Table \ref{time_comparison} shows the training times per epoch of several methods. Since the batch size is different on baselines, we test with fixed batch size to compare their time complexity fairly. The training epochs are obtained on the optimal batch-size settings reported in the papers. Since GF-CF requires no model training and SGL-ED is compared with LightGCN in terms of time complexity in the original literature, we ignore them. Among GCN baselines, LightGCN is the most efficient one, as it only keeps the core component (\textit{i.e.,} neighborhood aggregation) for training. On the other hand, the training time of GDE barely increases with the size of datasets, which is as fast as BPR and much faster than LightGCN; GDE-d takes more training time as Equations (\ref{smooth_capture}) and (\ref{rough_capture}) need to be repeated during each epoch of training. Overall, the running epochs of GDE, LightGCN, BPR are 120, 600, 450, and the whole running times are 502s (including preprocessing time), 5880s, 1480s, respectively; GDE has around 12x, 3x speed-up compared with LightGCN and BPR, respectively.

\subsection{Study of GDE (RQ3)}
 
\subsubsection{How Rough and Smoothed Features Contribute?} 
As shown in Figure \ref{rough_smooth}, the rough features behave like inherent features of the graph as it contributes to the accuracy without much model training and the training loss barely drops. On the other hand, the accuracy of the smoothed features significantly increases as the training proceeds, in the meanwhile the training loss drops sharply. Normally, we set $m_2\leq 0.1\times m_1$, since the accuracy is mainly contributed by the smoothed features while the rough features show a slight improvement to the accuracy. We also notice the improvement of rough features tends to be significant on dense datasets (\textit{e.g.,} MovieLens) and is hard to be identified on sparse datasets (\textit{e.g.,} CiteULike). A reasonable explanation is that the rough features can help emphasize the difference among nodes on dense datasets, where nodes are relatively connected to each other, whereas they are less important on sparse datasets as the nodes are already isolated on the graph.

\begin{table}[]
\caption{The accuracy of different designs to measure the importance of spectral features on CiteULike.}
\vspace{-0.2cm}
\scalebox{0.85}{
\begin{tabular}{c|c|c|ccc}
\hline
\multirow{2}{*}{Design} & \multirow{2}{*}{Function}      & \multirow{2}{*}{nDCG@10} & \multicolumn{3}{c}{Property}         \\ \cline{4-6} 
                        &                                &                          & Increasing & Pos Coef.  & Infinite   \\ \hline
\multirow{4}{*}{Static} & $\log(\alpha\lambda_t)$        & 0.1343                   & \checkmark & $\times$   & \checkmark \\
                        & $\sum_k\alpha_k\lambda_t^k$    & 0.1434                   & \checkmark & \checkmark & $\times$   \\
                        & $\frac{1}{1-\alpha\lambda_t}$  & 0.1464                   & \checkmark & \checkmark & \checkmark \\
                        & $e^{\beta\lambda_t} (\beta>0)$ & 0.1518                   & \checkmark & \checkmark & \checkmark \\
                        & $e^{\beta\lambda_t} (\beta<0)$ & 0.0322                   & $\times$   & $\times$   & \checkmark \\ \hline
Dynamic                 & Attention                  & 0.1296                   & \multicolumn{3}{c}{}                 \\ \hline
\end{tabular}}
\label{weight_func}
\vspace{-0.2cm}
\end{table}  

\subsubsection{What Kind of $\gamma(\cdot)$ Design Works Better? }
Table \ref{weight_func} lists some designs for weighting functions $\gamma(\cdot)$, and (1) Increasing, (2) Pos Coef, and (3) Infinite refer to: if it (1) is an increasing function, (2) has positive coefficients of Taylor series and (3) is infinitely differentiable, respectively. From the top to bottom, we set $\alpha=10$, $\alpha_k=1$, $\alpha=0.9$, $\beta=4$, $\beta=\mbox{-}2$. We can see the importance of the three properties is (1)$\gg$(2)>(3). The model shows poor performance when $\gamma(\cdot)$ is a decreasing function (\textit{i.e.,} the rougher features have higher importance), justifying our previous analysis that the smoother features are more important. Overall, the designs that satisfy all three properties outperform other designs. On the other hand, we notice that the dynamic design underperforms the static designs with an increasing function, which demonstrates that the dynamic design fails to learn the importance of different features, otherwise it should perform closely to above static designs. In addition, the running time comparison in Table \ref{time_comparison} also shows that a static design runs much faster than the dynamic design. Based on the above analysis and experimental results, we conclude that a static design is more effective and efficient for CF. 

\subsubsection{Effect of $\beta$ }
\begin{figure} \centering 
\vspace{-0.2cm}
\subfigure[CiteULike.] {  
\includegraphics[width=0.46\columnwidth]{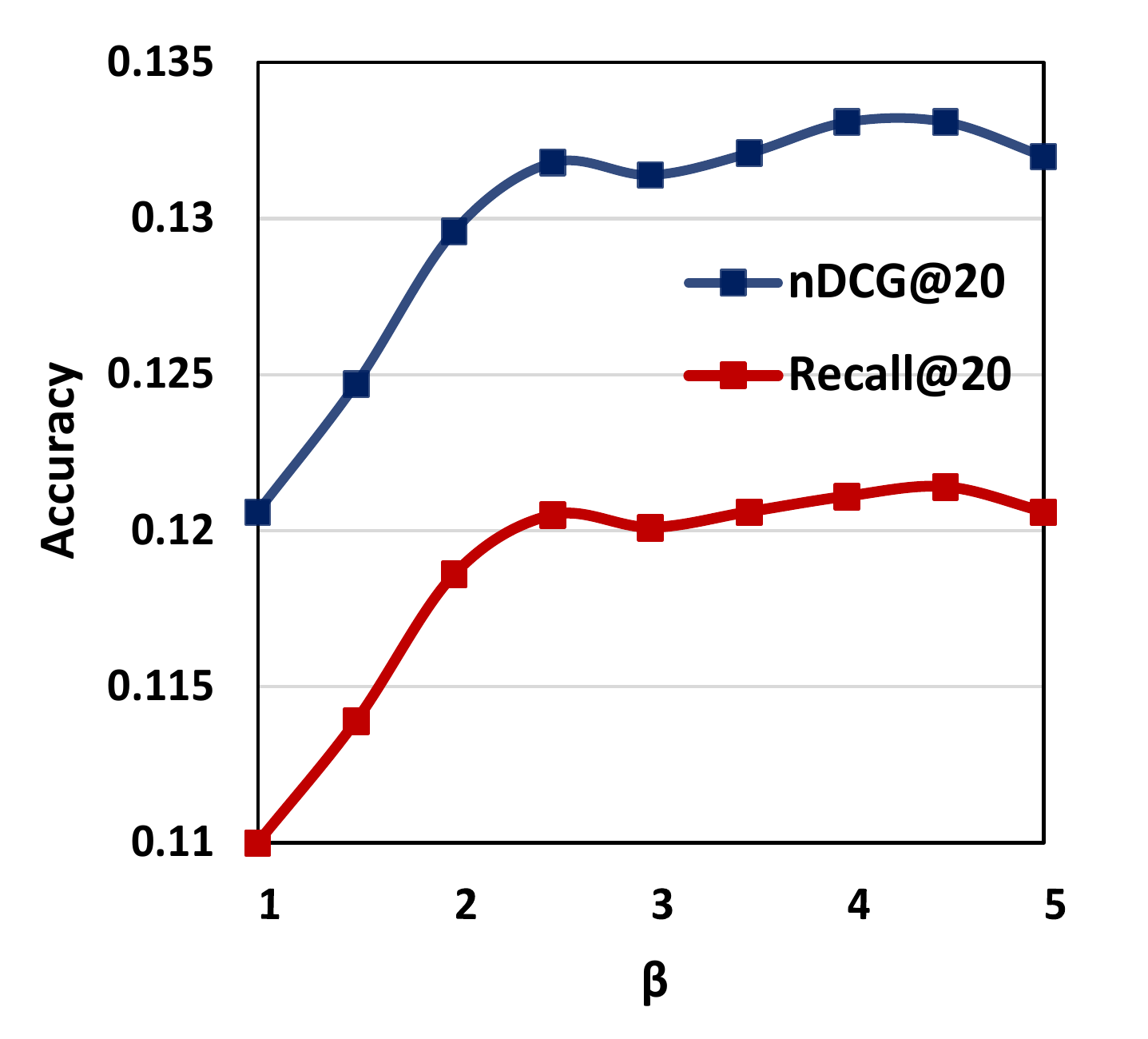} 
}  
\hspace{-0cm}
\subfigure[MovieLens-100K. ] {  
\includegraphics[width=0.46\columnwidth]{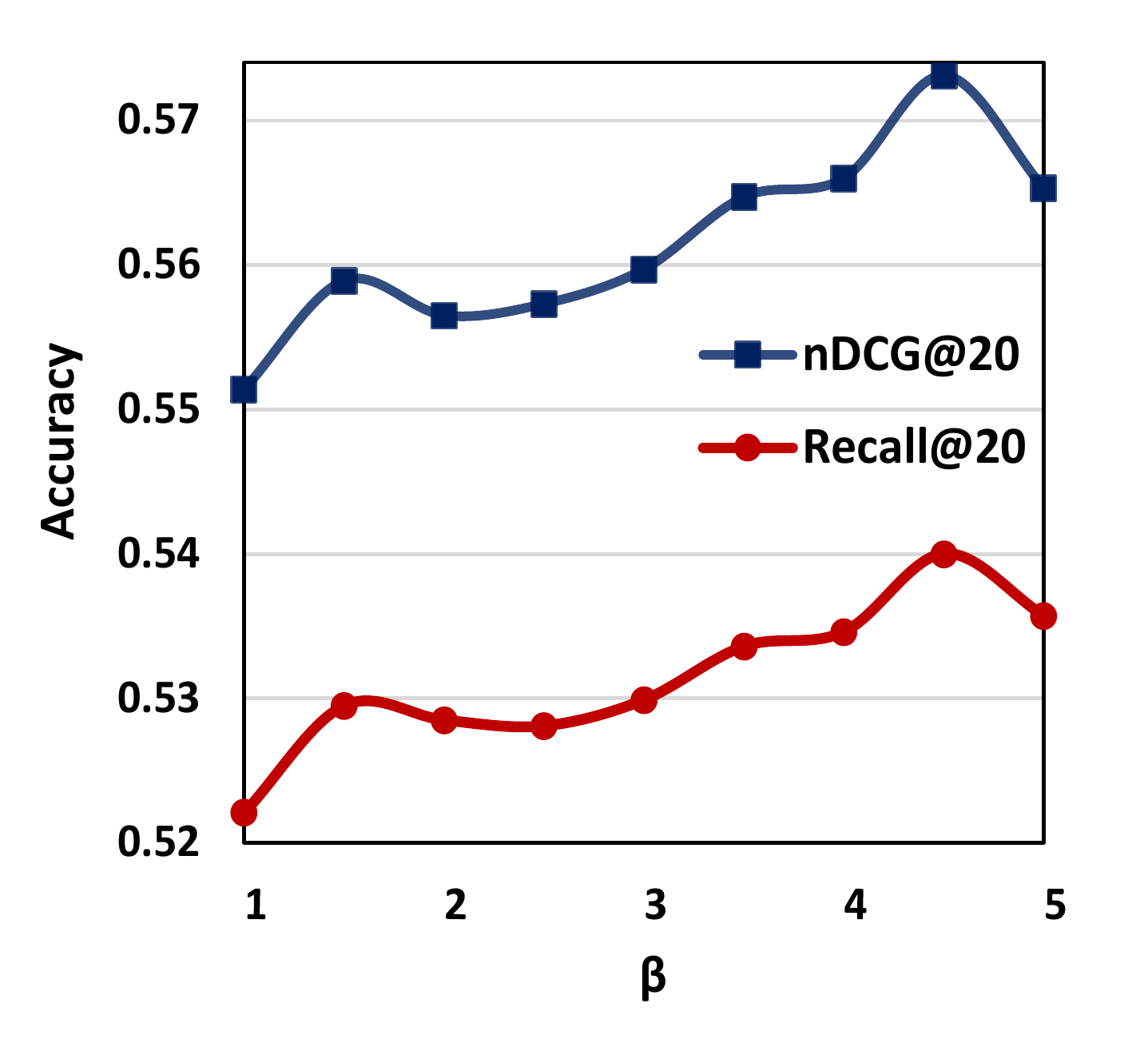} 
}
\vspace{-0.3cm}
\caption{Accuracy of GDE with varying $\beta$.}  
\label{beta}
\vspace{-0.3cm}
\end{figure}

Figure \ref{beta} shows the accuracy of GDE with varying $\beta$ on two datasets, where similar trends are observed on other datasets as well. We observe consistent improvements when the smooth features are emphasised ($\beta$ becomes larger). Particularly, the best accuracy is achieved at $\beta=4.5$ on both datasets, and the accuracy drops by 9.4\%, 3.5\% at $\beta=1$ compared with the best accuracy on CiteULike and 100K, respectively. We speculate the degradation is larger on CiteULike is because the smoothed features are more important on sparse data, on which users and items have fewer interactions and are more isolated on the graph.

\begin{table}[]
\caption{The accuracy of several GDE-variants evaluated by Recall@20.}
\vspace{-0.2cm}
\scalebox{0.86}{
\begin{tabular}{c|c|c|c|c|c}
\hline
                      & CiteULike       & Pinterest       & ML-1M       & Gowalla  &ML-100K         \\ \hline
plain               & 0.1146        & 0.1109        & 0.5303       & 0.1353   &0.5287        \\
drop               & 0.1198         & 0.1116        & 0.5390        & 0.1382  &0.5400          \\
drop+adaptive       & 0.1224       & 0.1147         & 0.5423       & 0.1449    &0.5275 \\ \hline
LightGCN            & 0.1066       &0.1069          & 0.5031       &0.1224     &0.5133 \\ \hline
\end{tabular}}
\label{effect_trick}
\vspace{-0.2cm}
\end{table}

\subsubsection{Ablation Study}
\begin{figure} \centering 
\vspace{-0.2cm}
\subfigure[GDE] {  
\includegraphics[width=0.46\columnwidth]{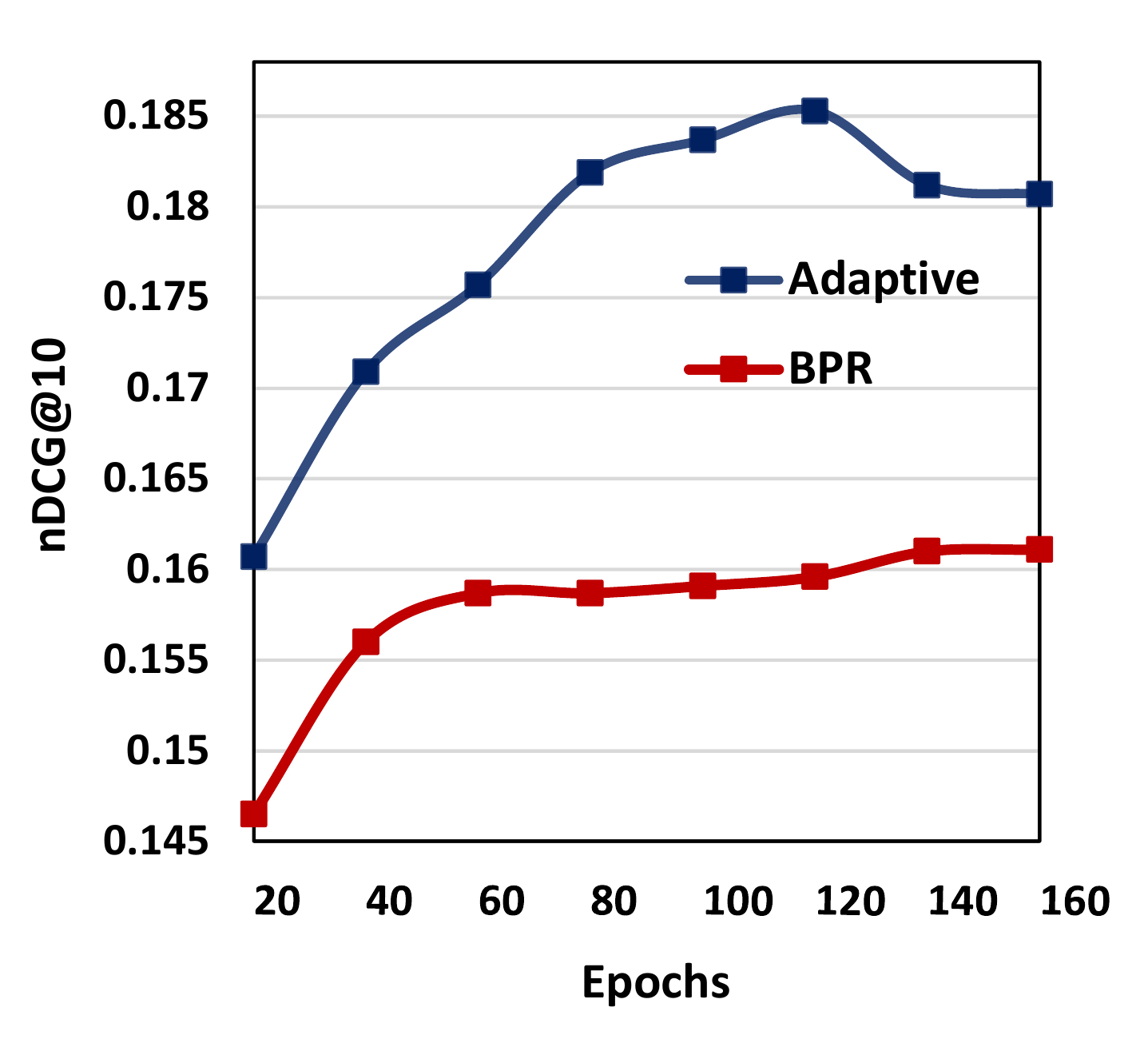} 
}  
\hspace{-0cm}
\subfigure[LightGCN ] {  
\includegraphics[width=0.46\columnwidth]{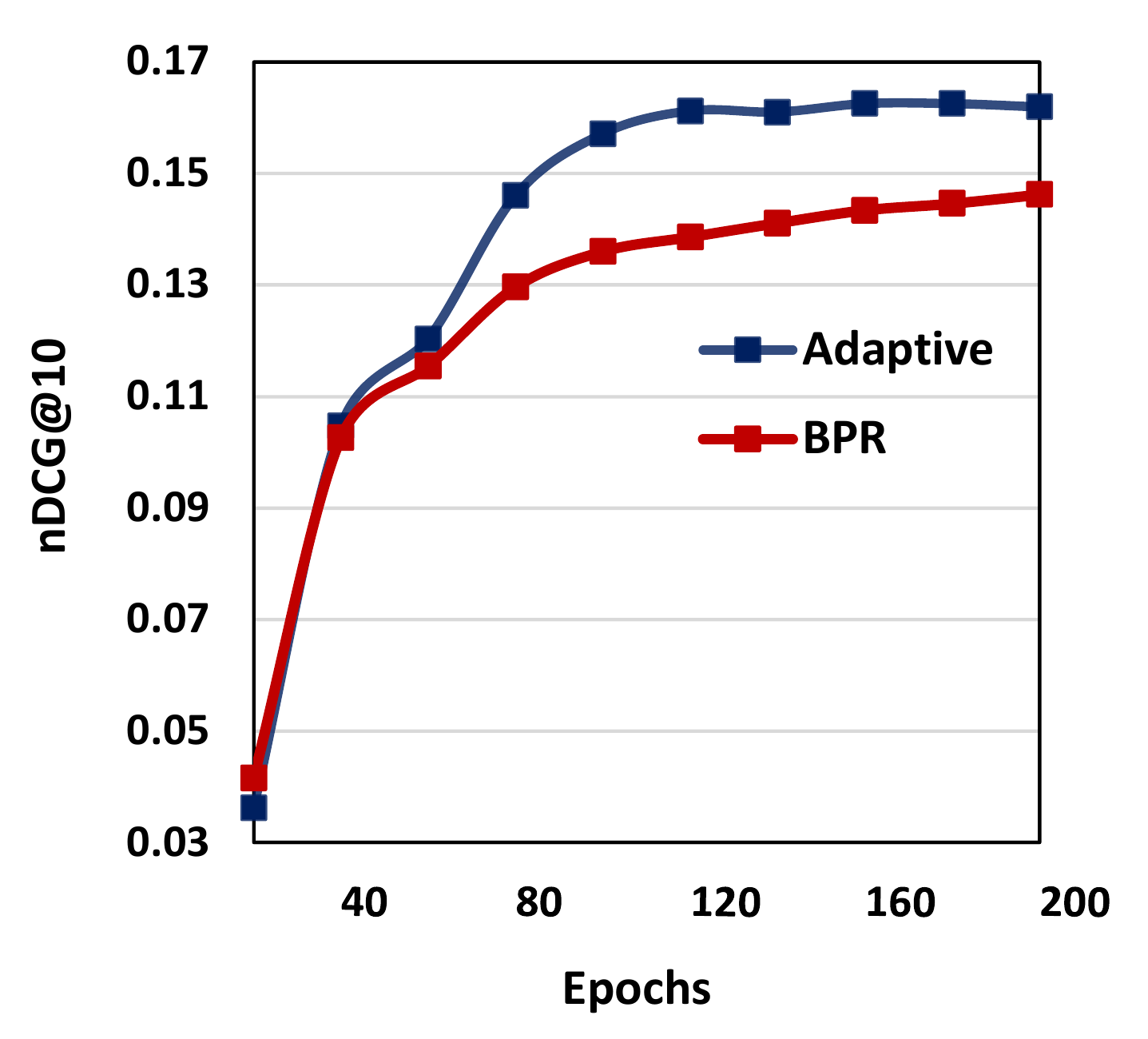} 
}
\vspace{-0.45cm}
\caption{The adaptive loss helps accelerate GCN training.}  
\label{adaptive_loss}
\vspace{-0.4cm}
\end{figure}

As shown in Table \ref{effect_trick}, we propose three variants: (1) with BPR loss and without dropout (plain), (2) with BPR loss and dropout (drop), (3) with adaptive loss and dropout (drop+adaptive). \\
\textbf{The effect of dropout.} Edge dropout is a commonly used technique to improve generalization on test data. We can identify the positive effect of it as a model with dropout shows improvements over the model without it across all datasets. In addition, we observe that the improvements on dense data (\textit{e.g.,} MovieLens 100K) tend to be more significant than sparse data (\textit{e.g.,} Pinterest).\\
\textbf{The effect of adaptive loss.} As shown in Figure \ref{adaptive_loss}, the adaptive loss largely reduces the training epochs of GCN-based methods including but not limited to GDE. From the reported results in Table \ref{effect_trick}, the adaptive loss also results in improvements on 4 out of 5 datasets. We speculate that the reason it reduces the accuracy on 100K is closely related to the data density; since we propose the adaptive loss to help alleviate the undersampling of the negative samples, where such issue is not serious on dense data as the negative samples are more likely to be sampled.  \\
\textbf{The effectiveness of GDE.}  It is fair to separate the improvement from the adaptive loss from overall improvements, as it can be applied on other GCN-based methods as well. We can see GDE with a common loss still outperforms competitive baselines including LightGCN, proving the effectiveness of our proposed designs.

\section{Related Work}
\subsection{Collaborative Filtering}
Collaborative Filtering (CF) is a commonly used technique in recommender systems. Early CF methods \cite{sarwar2001item} predict user tastes based on the taste of the users that share similar interests or the similar items to the item that the user has interacted with. Nowadays, most CF methods can be considered as extensions of the matrix factorization (MF) \cite{koren2009matrix} which characterizes users and item as embedding vectors and optimizes them based on direct user-item interactions. Claiming that estimating the rating simply based on a linear function is not enough to capture the complex user-item interactions, recent works resort to other advanced algorithms, such as neural networks \cite{he2017neural}, memory networks \cite{ebesu2018collaborative}, word embeddings \cite{liang2016factorization}, etc and propose more fine-grained architectures. Some works \cite{vinh2020hyperml,feng2020hme} explore the potential of learning in non-Euclidean space for recommender systems. Beyond optimizing based on <user, item> pairs, another type of methods \cite{donkers2017sequential,kang2018self} take the temporal factor into consideration by considering user actions as a sequence in chronological order to predict the future action.

\subsection{GCN-based CF Methods}
GCNs are developed from graph fourier transform for non-Euclidean data \cite{bruna2014spectral}. Kipf et al. \cite{kipf2017semi} having been the most commonly used GCN architecture in various fields including recommender systems, improves on one-layer of \cite{defferrard2016convolutional} and designs a multilayer architecture to hierarchically incorporate higher-order neighborhood. Most existing works adapt the vanilla GCN \cite{kipf2017semi} to recommendation such as rating-prediction \cite{berg2017graph,Zhang2020Inductive} and ranking \cite{wang2019neural} tasks. Much effort has also been devoted to tackle the limitations of GCNs for recommendation. Pinsage \cite{ying2018graph} defines a localized graph convolutions on the spatial domain for inductive learning and enjoys more flexibility. Non-linearity and transformation layers are shown ineffective for recommendation in recent works \cite{chen2020revisiting,he2020lightgcn}, their proposed simplified GCN architectures show less complexity and stable performance. In addition, some works bring other topics such as learning in hyperbolic space \cite{sun2021hgcf}, self-supervised learning \cite{wu2020self,yang2021enhanced} and so on to improve GCNs and achieve further success.\par  

Despite the superior performance achieved by the aforementioned methods, we notice that very limited works study how GCNs contribute to recommendation. LightGCN \cite{he2020lightgcn} empirically shows non-linearity and feature transformation are of no help in boosting accuracy, while the core component neighborhood aggregation has not been investigated. One recent work GF-CF \cite{shen2021powerful} theoretically shows LightGCN as well as some traditional CF methods are essentially low pass filters, sharing similarities with our analysis discussed in Section 3.1.2. However, our work differs from them in: (1) our analysis are more comprehensive as we focus on all spectral graph features and investigate how distinct spectral graph features contribute to CF rather than specific GCN designs, thus can be applied to graph-based recommendation including but not limited to GCN designs; (2) we demonstrate our proposed GDE significantly outperforms GF-CF under extreme data sparsity.

\section{Conclusion}
In this paper, we explored how GCNs facilitate recommendation and what graph information matters for recommendation from a spectral perspective. We especially showed how distinct spectral graph features contribute to the accuracy, found that only a very small portion of spectral features that emphasize the neighborhood smoothness and difference are truly helpful for recommendation. We then unveiled the effectiveness of GCNs by showing that stacking layers in GCNs emphasizes the smoothness. Based on the two important findings above, we pointed out the limitations of existing GCN-based CF methods and proposed a Graph Denoising Encoder (GDE) to replace neighborhood aggregation with a simple yet effective architecture. Finally, to tackle a slow convergence issue on GCN-based methods, we proposed an adaptive loss to dynamically adjust the gradients over negative samples, which accelerates the model training and results in further improvement. Extensive experiments conducted on five datasets not only demonstrate the effectiveness and efficiency of our proposed methods but also justifies our analysis and findings. We believe that the
insights of our work are inspirational to future developments of GCN architecture designs for recommender systems. In future work, we plan to analyze the potential of GCNs from other perspectives and apply our proposed models to other recommendation tasks.\\
\textbf{Acknowledgement.} We thank Tokyo ACM SIGIR Chapter committee members for their helpful and insightful comments.

\bibliographystyle{ACM-Reference-Format}
\bibliography{sample-base}

\end{document}